# RUSSIAN ACADEMY OF SCIENCES
National Geophysical Committee

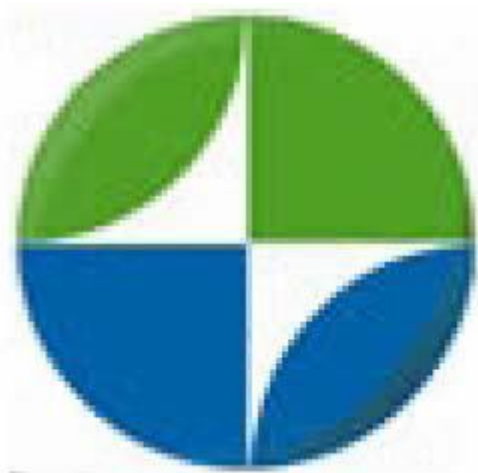

# NATIONAL REPORT
to the International Association of Geodesy
of the International Union of Geodesy and Geophysics
2015-2019
Presented to the XXVII General Assembly
of the International Union of Geodesy and Geophysics

2019
Moscow

**Abstract.** *In this National Report are given major results of researches conducted by Russian geodesists in 2015–2018 on the topics of the International Association of Geodesy (IAG) of the International Union of Geodesy and Geophysics (IUGG). This report is prepared by the Section of Geodesy of the National Geophysical Committee of Russia. In the report prepared for the XXVII General Assembly of IUGG (Canada, Montreal, 8–18 July 2019), the results of principal researches in geodesy, geodynamics, gravimetry, in the studies of geodetic reference frame creation and development, Earth's shape and gravity field, Earth's rotation, geodetic theory, its application and some other directions are briefly described. For some objective reasons not all results obtained by Russian scientists on the problems of geodesy are included in the report.*

The following institutes participated in the preparation of the Report:
[1] Far Eastern Federal University, Vladivostok, Russia
[2] Federal Scientific-Technical Center of Geodesy, Cartography and Spatial Data Infrastructure, Moscow, Russia
[3] Geophysical Center of the Russian Academy of Sciences, Moscow, Russia
[4] Geophysical Survey of the Russian Academy of Sciences, Obninsk, Russia
[5] Institute of Applied Mathematics, FEB RAS, Vladivostok, Russia
[6] JSC Institute Hydroproject, Moscow, Russia
[7] National Research Institute for Physical-Technical and Radio Engineering Measurements (VNIIFTRI), Mendeleevo, Moscow Reg., Russia
[8] Moscow State University of Geodesy and Cartography, Moscow, Russia
[9] Pulkovo Observatory, Saint Petersburg, Russia
[10] Russian society of geodesy, cartography and land management, Moscow, Russia
[11] Siberian State University of Geosystems and Technologies, Novosibirsk, Russia

The Report was prepared by the following scientists:
[1,5] Gerasimenko M., [9] Gorshkov V., [3] Kaftan V., [11] Kosarev N., [9] Malkin Z., [11] Mazurov B., [7] Pasynok S., [10] Pobedinsky G., [2] Popadiev V., [8] Savinykh V., [2] Sermiagin R., [1,5] Shestakov N., [4] Steblov G., [8] Sugaipova L., [6] Ustinov A.

# Content

# Preface

The National Report contains information on the achievements of Russian researchers in the field of geodesy for the period 2015-2019. The information is collected from published materials and is a collection of short abstracts on selected scientific publications. It briefly describes the results of principal research in geodesy, geodynamics, gravimetry, in the studies of geodetic reference frame creation and development, the Earth's shape and gravity field, the Earth's rotation, geodetic theory, its application and some conterminous areas of research. The information is presented on aggregated headings, generalized to the relevant topics of research on the structural content of the IAG / IUGG. The book also provides information on international research in the field of geodesy with the participation of Russian scientists.

The authors of the collection are members of the Geodesy Section of the National Geophysical Committee of the Russian Academy of Sciences. The collection contains far from complete information about the entire volume of research in the field of geodesy. The authors hope that they were able to summarize the most significant results and bring them to the international scientific community.

The more principal studies are listed below.

The development of the Fundamental Astro-Geodetic Network (FAGN) and subsequent levels of the state geodetic network are of great importance for the implementation of the modern national geocentric coordinate system.

Substantial improvement of the national and international coordinate frame is provided by the creation of new high-precision electro-optical measuring tools of the laser systems of the new generation "Tochka" and the network of laser ranging of the Earth satellites.

International studies on the improvement of the reference catalogs of the latest realisations of the International Celestial Reference Frame (ICRS) as well as on the relationship between the ICRF and Gaia-CRF contribute to the effective development of the unite complex of global coordinate support. In the frame of international cooperation various approaches are tested in different analysis centers to extend the IERS strategy on the rigorous TRF+CRF+EOP combination.

Important theoretical studies have been carried out in the field of approximation of the gepotential. The most commonly used scaling functions, wavelets and radial basis functions are described and realised in the study. Recommendations for joint processing of heterogeneous measurements by the least squares method are preseted in respective monograph.

An effective algorithm has been developed to account for ionospheric effects in carrier-phase GNSS measurements.

***Section of geodesy of the National Geophysical Committee***
*Dr. V.P. Savinykh, Chairman, Corr. Member of RAS*
*Dr. V.I. Kaftan, Vice-chairman*

# Reference Frames

**Kaftan V.[1], Malkin Z.[2], Pasynok S.[3], Pobedinsky G.[4], Popadiev V.[5]**

[1]Geophysical Center of the Russian Academy of Sciences, Moscow, Russia
[2]Pulkovo Observatory, Saint Petersburg, Russia
[3]National Research Institute for Physical-Technical and Radio Engineering Measurements (VNIIFTRI), Mendeleevo, Moscow Reg., Russia
[4]Russian society of geodesy, cartography and land management, Moscow, Russia
[5]Federal Scientific-Technical Center of Geodesy, Cartography and Spatial Data Infrastructure, Moscow, Russia

The 2015 is the year of the 20th anniversary of the Concept of transition of the topographic and geodetic production of the Russian Federation to the autonomous satellite methods of coordinate determination developed in 1995 Central "Order of Honor" Scientific Research Institute of Geodesy, Aerial Survey and Cartography named after F. N. Krasovsky (TSNIIGAiK) of the former Federal Service of Geodesy and Cartography of Russia.

The geodetic fundamental of the Concept designed to solve the problem of coordinate support of the country in connection with the advent of satellite era of coordinate determinations. It was standardized the transfer of the national geodetic network in a single geocentric coordinate system based on the use of GNSS technologies. Subsequent normative-technical documents (geodetic, mapping instructions, rules and regulations) and national standards, the hierarchical structure of the satellite geodetic network, include the following network levels:

- network of permanent GNSS observations is a Fundamental Astro-Geodetic Network (FAGN);

- network of periodic GNSS observations — Precise Geodetic Network (PGN);

- the network once detrmined points - a Satellite Geodetic Network of 1 class (SGN-1).

The practical implementation of the Concept began in 1995-1996. Now it is ongoing. Geodetic satellite network of the 38 points of the FAGN and about 250 points of PGN were built and subject to the adjustment and cataloging to 2009 [Басманов и др., 2015].

By January 1, 2015 the state geodetic satellite network consisted of 4,624 points. The FAGN consisted of 54 stations, 45 of which are permanent and 9 are periodically determined. Part of the points the FAGN consisted of 13 RAS stations, 5 Rosstandart stations and 36 Rosreestr stations, 3 points are collocated with the VLBI stations, 8 points are joined with the stations of differential correction and monitoring of the Russian Space Agency. The PGN consisted of 326 points and SGN-1 consisted of 4 244 points [Басманов и др., 2015].

The state geodetic satellite network of the Russian Federation includes 67 points of FAGS, 363 points of VGS and more than 5 000 points of SGS-1 at the beginning of 2019. Four new FAGN stations were installed in 2018, including the northernmost FAGN station "Barentsburg" on the Svalbard Archipelago, the appearance of which is shown in Fig. 1.

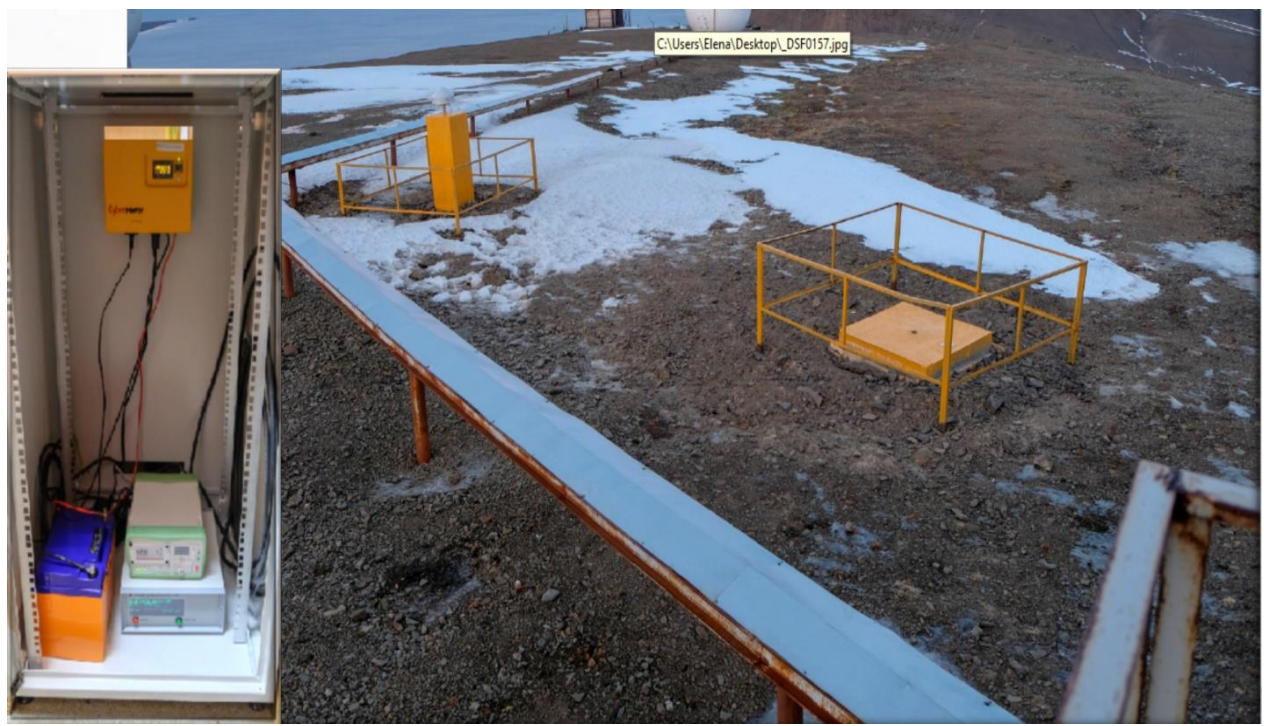

Fig. 1. FAGN station "Barentsburg".

The information about the FAGN points the location and their coordinates in the local coordinate system presented on the website of the "Centre of geodesy, cartography and SDI" http://cgkipd.ru/opendata/geodesy/fags_localcs. The scheme the FAGN stations is shown at Fig. 2.

The permanent stations are stationary astro-geodetic observatories equipped by a complex of precision apparatuses: frequency standards, meteorological sensors, equipment for local earth's surface deformation monitoring in the area of the observatory and the stability of the position of the antenna pillars, etc. All stations are fundamentally earth fixed to ensure long-term stability of their position both in plan and height.

The development and the establishment of the geodetic coordinate system of 2011 (GSK-2011) is a logical step in the development of geodetic maintenance in Russia.

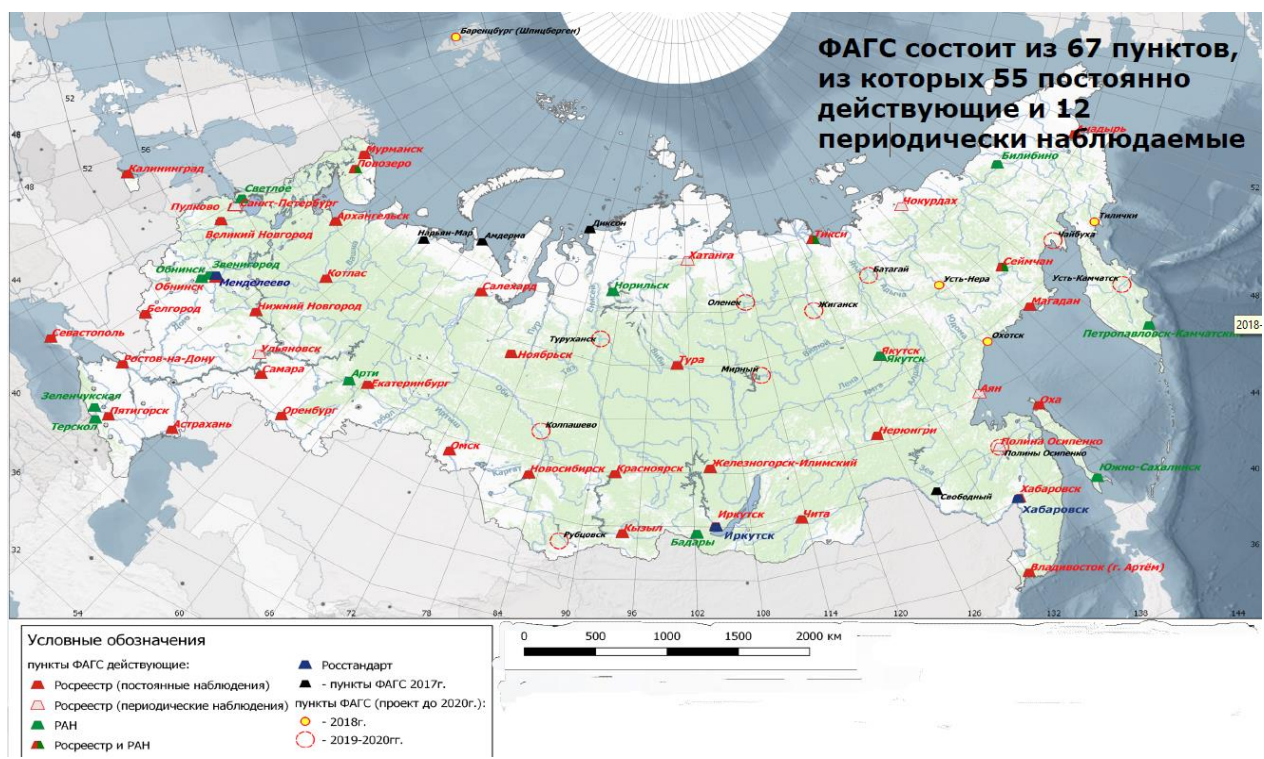

Fig. 2. FAGN stations location.

State geodetic coordinate system of the Russian Federation GSK-2011 is a geocentric coordinate system. According to the principles of the fixation in the body of the Earth, GSK-2011 are identical to the International Terrestrial Reference System (ITRS) established in accordance with the recommendations of the International Earth Rotation Service (International Earth Rotation and Reference Systems Service — IERS).

The basis of the coordinate system GSK-2011 was the state satellite geodetic network FAGS. The overall composition of the sites included in the Astro-Geodetic Network adjustment is characterized by the following data:

Total number of items included in the adjustment 46;

Among them:

– Russian 38;

– Overseas 8;

Number of initial points 21;

Including:

– Russian 13;

– Overseas 8;

The initial data for the adjustment includes:

• files daily satellite observations at the FAGN stations in the interval of 2 years (2010-2011);

• files of satellite coordinates;

• simulated ionospheric delay files;

• file details about the failures of the satellite equipment GPS/GLONASS;

• Earth rotation parameters.

Processing of daily measurement sessions was performed using BERNESE 5.0 software. The traditional scheme of calculations on double differences of phase measurements including the following main stages of calculations was used:

1. Numerical integration of the motions of satellites GPS/GLONASS for calculation of their coordinates on the points of observations;

2. Transformation and preliminary control of observational data;

3. Transformation of the initial coordinates of the stations to the era of the observation session;

4. Calculation of amendments hours receivers on code measurements;

5. The formation of the single difference phase measurements, advanced control and rejection of gross values;

6. Resolution of phase ambiguities (calculations on individual baselines);

7. Joint adjustment of the entire network as «free»;

8. Calculation of transformation Helmert parameters between the calculated coordinates of the reference points and the epoch of observation values;

9. Control of deviations of the calculated coordinates of the reference points and in the case of rough results — the exclusion of the point from the reference and the implementation of a new network adjustment;

10. Transform all calculated coordinates of a "free" network using the Helmert parameters and the accuracy estimation.

The initial coordinate values of the points used in step 3 were selected from the tables of their values and the displacement velocities for the 2011.0 epoch. The IGS point coordinates were selected from the international catalogue, and for other points a priori values of coordinates and velocities calculated using the geodynamic model NUVELL.

The mechanism of tying FAGN to a "skeleton" IGS reference points network (steps 8-10) based on IERS widely used method of "soft coordination". According to this method, when equating the daily measurement sessions at the first stage, the network is considered as "free", i.e. the coordinates of the control points are not fixed rigidly, but are calculated together with the coordinates of other points. At the second stage, the Helmert transformation of the calculated "free" network is carried out under the condition of the minimum sum of the squares of the deviations of the obtained coordinates of the control points from their accepted values.

Root mean square errors (RMS) of adjusted FAGN coordinates amounted to 0.1–1.0 cm-up (point FAGN "Vladivostok" — 1.9 cm) and 0.2–1.5 cm in height (point FAGN Vladivostok — 2.14 cm) [Басманов и др., 2015].

The history of the state geodetic satellite network and the state geodetic coordinate system GSK-2011 is described in detail in the monograph "GLONASS and geodesy", prepared under the leadership of the outstanding Soviet and Russian scientist Gleb Demyanov [GLONASS, 2016].

Geodetic coordinate system GSK-2011 was established by the Decree of the Government of the Russian Federation on December 28, 2012 № 1463 "On unified state coordinate systems" for use in the performance of geodetic and cartographic works. After the adoption of the new Federal Law of December 30, 2015 № 431-FZ "On geodesy, cartography and spatial data and on amendments to certain legislative acts of the Russian Federation" this coordinate system was re-established without changes by the decree of The Government of the Russian Federation of November 24, 2016. No. 1240 "On the establishment of state coordinate systems, the state system of heights and the state gravimetric system".

Reference materials on GSK-2011 and related issues are presented in the last section of the scientific and technical compendium "Astronomy, Geodesy and Geophysics" [Попадьёв и др., 2018].

The study [Бовшин, 2019] deals with a high-precision geodetic network densification by means of GNSS based geodetic solutions, in view of the fact that initial data are represented in different reference frames. Indeed, reference station positions are represented in GSK-2011 terrestrial reference frame whereas GNSS satellite ephemeris are represented in another reference frames, such as ITRFs, WGS84, etc. Two methods are considered in the paper to provide GNSS observations with a correct processing procedure: in the first method all initial data are translated into the satellite ephemerid reference frame, and in the second one - into GSK-2011 terrestrial reference frame. To supply these translations, 7-parameter Helmert time-dependent transformations for both methods were developed. After all, GNSS observations over period of 2010-2016 of the permanent stations of Moscow region were processed with the second method and some processing results were displayed in the paper to show an efficiency of the developed techniques.

"Quick Guide to work in the geodetic system of coordinates of 2011" is prepared by Popadiev V.V. The "Explanation to the geodetic coordinate system of 2011" is constantly updated, developed by G. N. Efimov, V.I. Zubinsky, V.V. Popadiev.

Electronic catalogues of geodetic points that have been transferred to territorial administration of Federal registration service composed with the aim of bringing the data about the points of the state geodetic network in the GSK-2011 to the consumers on the whole territory of the Russian Federation. To ensure continuity with the materials created in the coordinate systems SK-42 and SK-95, catalogs of triangulation, trilateration and polygonometry points of class 1 - 4 in the coordinate system GSK-2011 were compiled on the nomenclature sheets of the state topographic maps of scale 1:200 000 [Gorobets et al., 2015]

In order to provide consumers with accurate ephemerides of GNSS I satellites in TSNIIGAiK as the leading organization of the Russian Federation in the field of geodes the international ephemeris center of GNSS was established in 2004 in accordance with the agreement of the Federal service of geodesy and cartography of Russia (Roskartografiya) and the Federal Agency of cartography and geodesy of Germany (BGK). Currently the Precise Ephemeris Center is supported by the Federal scientific and technical center for geodesy, cartography and spatial data infrastructure. On the official website of the Precise Ephemeris Center (http://rgs-centre.ru/) presents:

- GNSS observation data at the FAGN points;

- additional information about the FAGN points;

- precise ephemeris of GLONASS satellites;

- coordinates and rates of change of the coordinates of the FAGN points in the coordinate system of GSK–2011.

The solution of issues of further development of the state coordinate system is primarily associated with the improvement of the network of FAGN points as the base of the coordinate system.

Design of new permanent stations should be conducted with consideration of tectonic structure of territory of Russia and the potential for transmission of observations in a single center in the real time. On the other hand, an important requirement for the location of FAGN points is their relatively uniform distribution on the territory of Russia with an average distance of about 500-800 km between points. These requirements for the placement of new fags points, firstly, will provide a differentiated approach to determining the rates of change of coordinates in time for different geotectonic structures, and secondly, will provide more favorable conditions for the spread of a single coordinate system and velocities to the points of geodetic networks of lower level (primarily with additional or periodic determinations of points) and thirdly, will create more favorable conditions for the development of systems of functional additions to GNSS (RTK, VRS, PPP, etc.) [Переход ... , 2016].

One of the mass technologies of geodetic support of consumers at the present time and in the future will be the transmission over the Internet of measuring and correcting information from the points of FAGN and differential geodetic stations to determine the coordinates.

In accordance with the current legislation of the geodetic network of differential stations to provide geodetic works has the right to establish the physical and legal entities, bodies of state power and bodies of local self-government. Examples of networks of differential geodetic stations are:

- system NAVIGATIONLINK ensuring Moscow (SNGO Moscow), state unitary enterprise Mosgorgeotrest (http://sngo.mggt.ru);

— the network of permanent differential stations GSI, JSC "GEOSTROYIZYSKANIYA" (http://topnet.gsi.ru);

— satellite geodetic network of base (reference) stations on the territory of St. Petersburg and Leningrad region "GEO - SPIDER", LLC " NPP "GEOMATIC" (http://geospider.ru).

In addition to the owners of networks of differential geodetic stations services for the provision of measurement and correction information provided by organizations that are not the owners of these networks.

Among such services currently operated in the Russian Federation are the following:

— HIVE, NPK "Industrial geodetic systems" (https://hive.geosystems.aero);

— SmartNet Russia, ООО "navgeokom" (http://smartnet-ru.com);

— National Network of high-precision positioning, NP operators of high-precision satellite positioning networks (http://nposvsp.ru) [Pobedinsky, 2016, Pobedinsky, 2018].

Placement of differential geodetic stations and points of geodetic networks of special purpose, reports on the creation of which are placed in the Federal Fund of spatial data, is given on the website of the "Center of geodesy, cartography and spatial data infrastructure" at http://cgkipd.ru/opendata/GSSN/. An example of the location of differential geodetic stations in St. Petersburg is shown in Fig. 3.

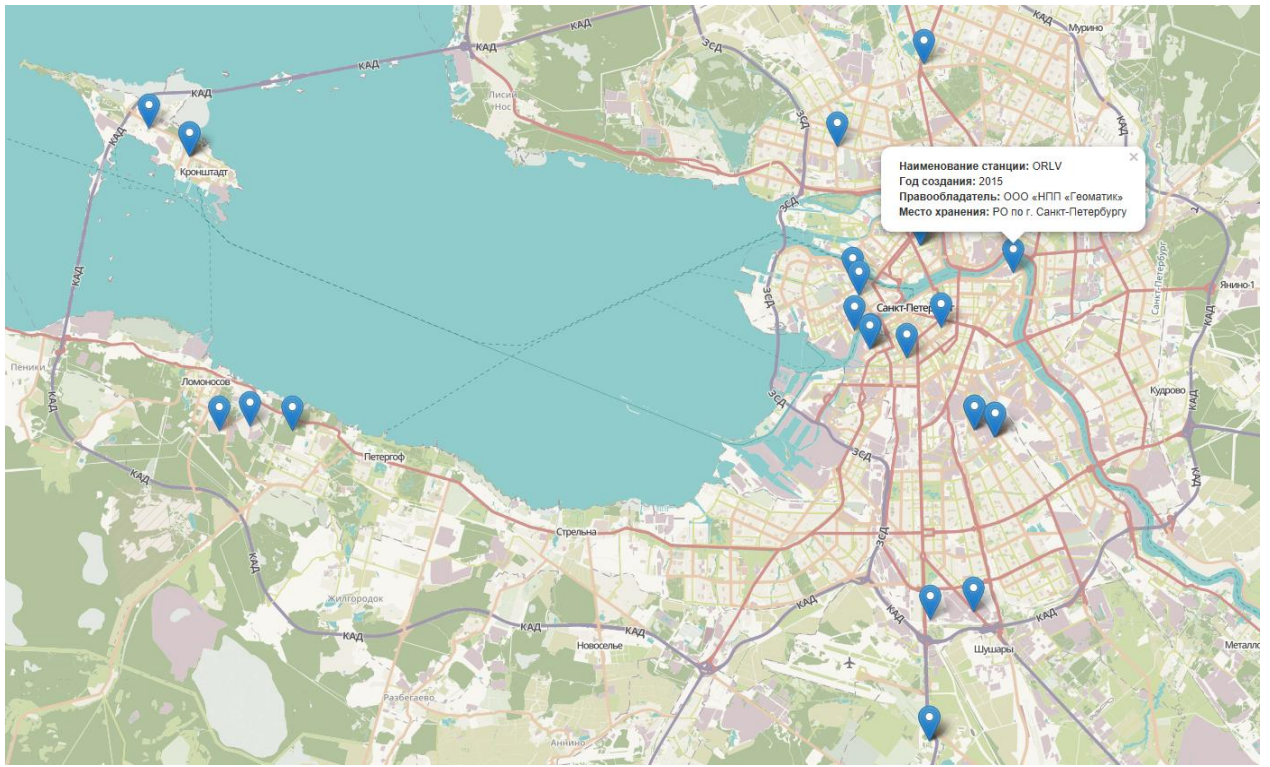

Fig. 3. Location of differential geodetic stations in St. Petersburg.

The Main Russian Vertical Reference Frame (MRVRF) lines of Russia form a uniform network of level circuits consisting of ~ 170 closed loops of 1st order and about 1000 polygons of class 2nd order and mixed polygons (See fig. 4).

The perimeters of the 1st order circuits are from 190 km to 2.6 thousand km (an average of 980 km) in the European part of Russia; from 400 km to 4.7 thousand km (an average of 2.2 thousand km) for the Siberia and the Far East. The total length of the MRVRF leveling lines in Russia is 325 000 km, 155 000 km of which is the 1st order, 170 000 km is the 2nd order. The average measurement epoch in the MRVRF of Russia corresponds to 1983, for the 1st order lines - 1989, and for the 2nd order lines of 1977. When processing high-precision leveling, the possibility of accurately calculating the correction for the transition from the sum of measured elevations to the height difference using the global gravity model is shown in [Попадьёв, Кулиев, 2017].

Data of GNSS observation at the Russian continuously operated station and GLONASS ephemerides are accessible from https://rgs-centre.ru/.

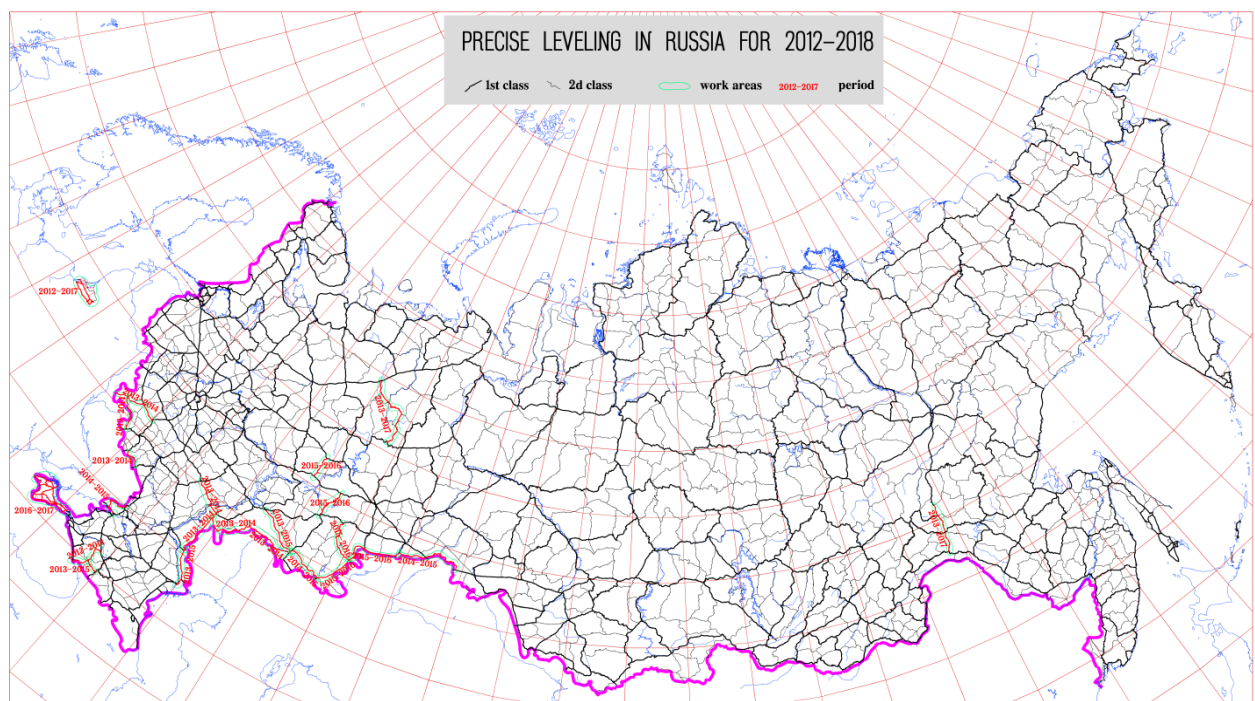

Fig. 4. Main Russian Vertical Reference Frame (MRVRF)

The third realization of the International Celestial Reference Frame, ICRF3, has been developed by the ad hoc IAU Working Group (Chairs C. Jacobs and P. Charlot) with active participation of the Geodesy Section members [Malkin et al., 2015]. The final report on ICRF2 was presented at the IAU General Assembly 2018, and the Assembly approved ICRF3 as official IAU celestial frame starting from 1 January 2019. This realization replaced previous frame ICRF2 [Fey et al., 2015].

Methods of computation of the combined catalogs of the extragalactic radio source positions were developed [Лопез, Малкин, 2018]. Adopted procedure of the construction of combined catalog consists of two stages. First, the difference between the input catalogs and the celestial reference system ICRF (ICRF2) is presented in a series of spherical functions using the Brosche method. Each catalog is then corrected for the found differences (catalog system) and the corrected catalogs are averaged. Formed catalog is called PC1. Then the input catalog systems obtained at the first stage are averaged, and the resulting average system is considered as systematic correction to the ICRF catalog. The addition of the average system to the catalog obtained at the first stage results in a final combined catalog PC2. Figures 5 and 6 show comparison of combined catalogs with ICRF2.

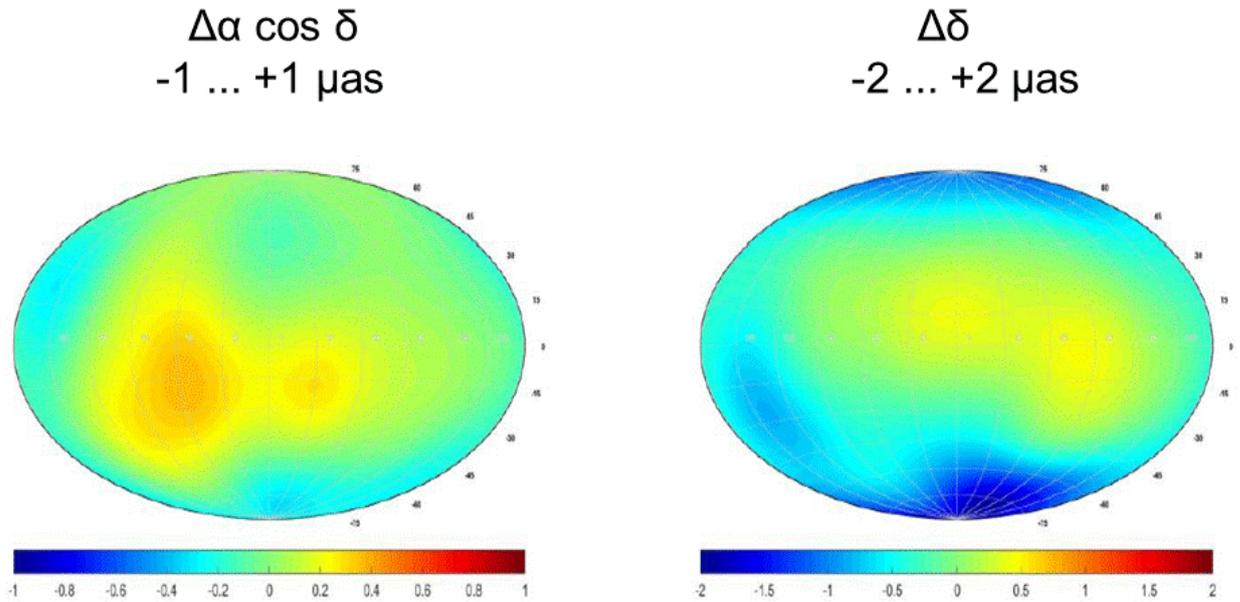

Fig. 5. Differences between catalogs PC1 and ICRF2.

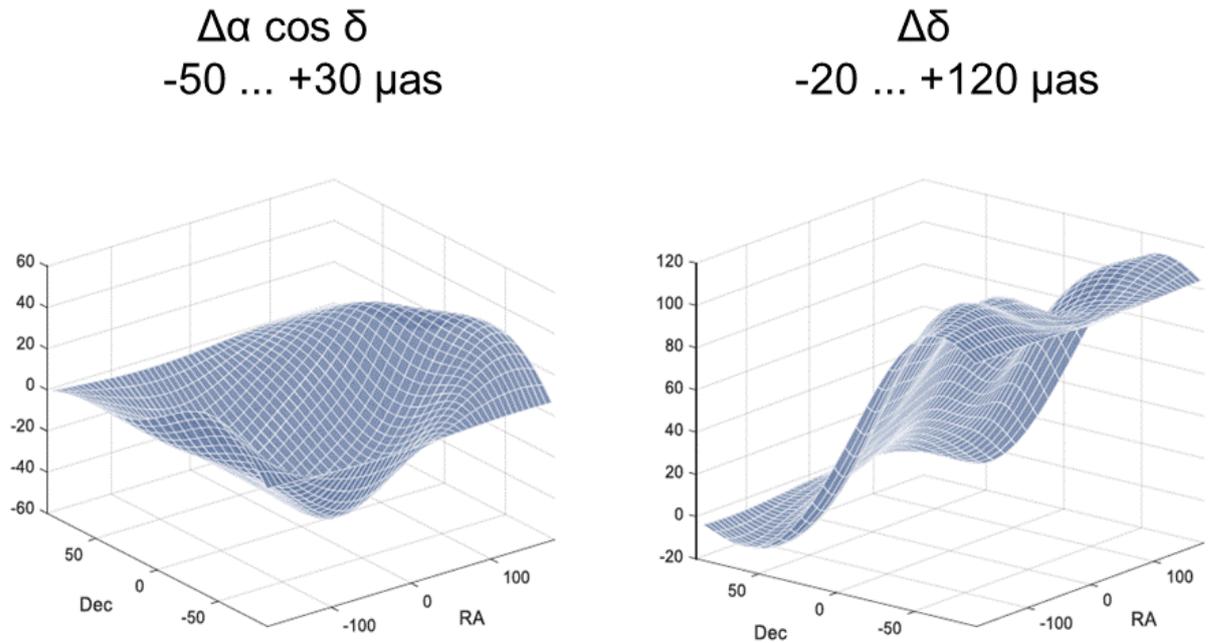

Fig. 6. Differences between catalogs PC2 and ICRF2.

Impact of correlation information on the orientation parameters between two celestial reference frames was studied [Sokolova, Malkin, 2016]. Three methods of accounting for the correlation information was compared: using the position errors only, using the RA/DE correlations reported in radio source position catalogues in the IERS format and using the full correlation matrix. Test computations were performed with nine catalogues computed in 8 VLBI analysis centers. The analysis results showed that using the RA/DE correlations only slightly influences the computed rotational angles, whereas using the full correlation matrices leads to substantial change in the orientation parameters between the compared catalogues.

It is also possible that accounting for a full correlation matrix will be essential not only at definition of mutual orientation, but also at decomposition of coordinate's differences by orthogonal functions.

In view of newest results of the Gaia mission, the problem of connection of the optical and VLBI frames becomes very essential. Several studies were devoted to this topic [Малкин, 2016a; Malkin, 2018a; Malkin, 2016b]. The following aspects were considered: increasing number of optically bright radio sources with highly accurate VLBI positions, dependence of the orientation parameters on the optical magnitude limit of the sources in common between two frames, using radio stars, increasing the number of southern radio sources with accurate positions. Different possibilities to improve the link accuracy between ICRF and Gaia-CRF were discussed. To simplify a selection of optically bright sources for intensive VLBI observations, OCARS (Optical Characteristics of Astrometric Radio Sources) catalog was updated [Malkin, 2018b].

One of the systematic errors affecting the accuracy of the link between optical and radio frames is the Galactic aberration (Fig. 7).

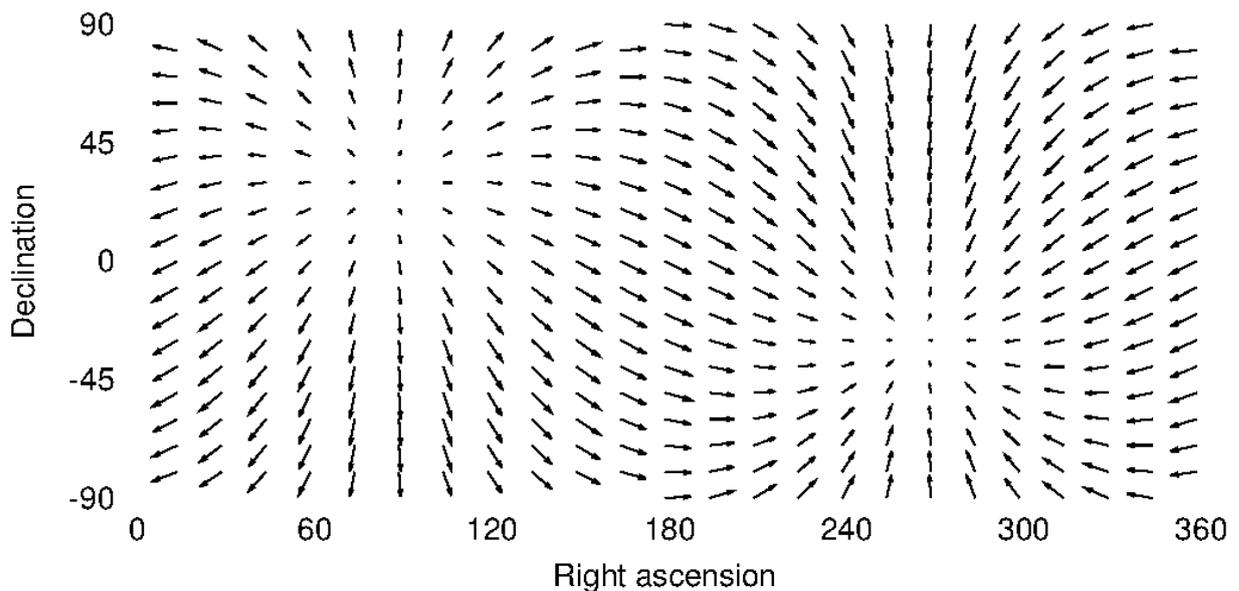

Fig. 7. Impact of the Galactic aberration on the source coordinates. Maximal arrow has the length of 5 μas.

It should consistently accounted for in both ICRF and Gaia catalogs.

Analysis of the impact of Galactic aberration on the orientation parameters between two frames were discussed [Malkin, 2015; Малкин, 2016d]. Preliminary results showed that this impact is at a level of about one microarcsecond.

Impact of quasar proper motions on the alignment between the International Celestial Reference Frame and the Gaia reference frame was studied in [Liu et al., 2018]. Several VLBI and Gaia source position time series were compared. Comparison of the radio optical source motions for about 600 common sources showed difference in the rotation and glide components of 2−4 µas/yr. Therefore, proper motions should be considered for highly accurate ICRF–Gaia link.

Impact of the radio source structure on the computed radio source position is one of the important problem in improving the accuracy of VLBI results. A new approach to solution of this problem based on four-parameter two-component model was proposed in [Titov et al., 2017]. It is based on the processing geodetic data without using a special observing programs. Implementation of this model to analysis of the CONT14 data showed the displacement of the declination of radio source 0014+813 of 0.070 mas with respect to its reference position.

One of the most important problems discussed in the literature and, in particular, in the IAG Sub-Commission 1.4 is the consistency between the TRF, CRF, and EOP. Different aspects of this topic were considered in [Malkin, 2105; Malkin et al., 2016; Malkin et al., 2017]. While ICRF is derived from VLBI only data processing, ITRF is a product of joint processing of data obtained from four space geodetic techniques: GPS, SLR, VLBI, and DORIS. Currently the procedure of computation of the ITRF provide rigorous TRF+EOP combination. Various approaches are tested in different analysis centers to extend the IERS strategy on the rigorous TRF+CRF+EOP combination.

VNIIFTRI as the Russian Main Metrological Center of Time, Frequencies and Earth Rotation Service carried out the rapid EOP processing based on GNSS, VLBI and SLR observations for many years. At present, VNIIFTRI supports and improves 38 State standards, 19 secondary standards, 23 rigs of highest accuracy, over 120 working standards and calibration rigs for various fields of measurement. VNIIFTRI performs the duties of the Main metrological center of the State service of time, frequency and the Earth rotation parameters determination (SSTF).The East-Siberian branch of FSUE «VNIIFTRI» is an autonomous structural subdivision of FSUE «VNIIFTRI» and acts in accordance with The Rules of FSUE «VNIIFTRI», The Branch Regulations and Russian legal system. The major aim of foundation of the East-Siberian branch is carrying out of technical-scientific activity of measurement assurance either in the territory of Eastern Siberia or the whole country [Ignatenko, Donchenko, Blinov, 2018].

The observation sites are situated in Mendeleevo (Moscow reg.), Novosibirsk, Irkutsk, Khabarovsk and Petropavlovsk-Kamchatsky. Three of these sites (mdvj, novm and irkj) are fundamental sites of International Terrestial Reference Frame (ITRF). Two of these sites (mdvj and irkj) are the sites of the particular collocation (GNSS and SLR) and the laser systems of new generation «Tochka» are on these sites. All five sites are included in Russian fundamental astronomical-geodetic network (FAGS). Site mdvj also is site of EUREF Permanent GNSS Network (EPN).

Laser ranging measurements FSUE "VNIIFTRI" is held since the 1970 (Fig. 8). Now VNIIFTRI and its East-Siberian branch are equipped the modern SLR systems (Fig. 9). These stations have the similar equipment:

- the laser location system (Russian designed);

- time and frequency standards (H-masers) and synchronization system;

- precise gravimeters;

- GPS/GLONASS receivers;

- Local Geodetic Network;

- Communication Network.

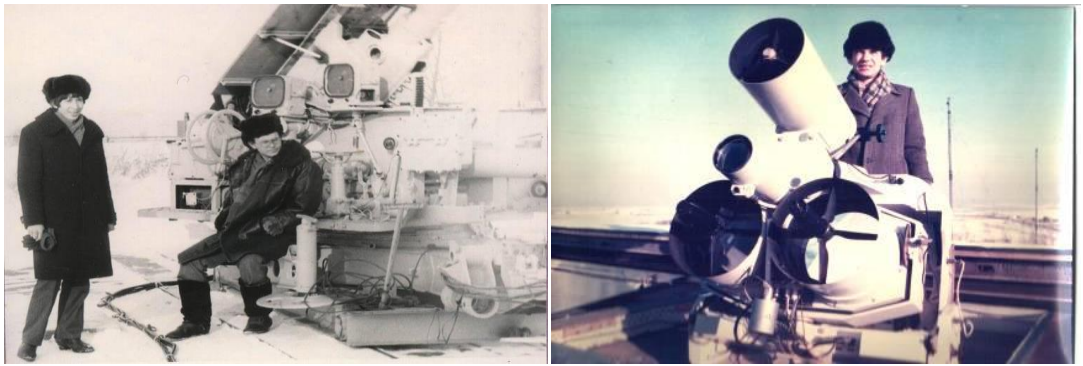

Fig. 8. SLR in VNIIFTRI in the past.

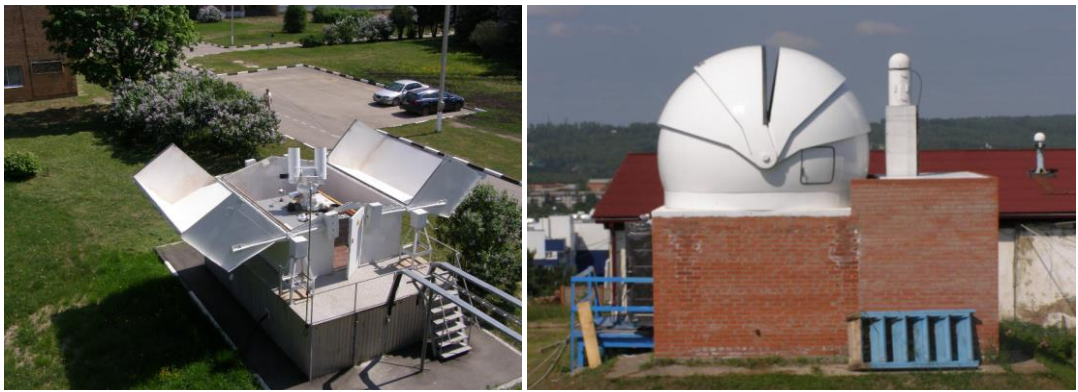

Fig. 9. SLR in VNIIFTRI and its East-Siberian branch now.

The accuracy of these instruments is on centimeter level. SLR stations are used together with standards time, frequency, national time scale and length [Игнатенко И.Ю. и др.]:

- Sate time and frequency standard in Mendeleevo UTC(SU);

- State standard of length in Mendeleevo;

- Secondary time and frequency standard in Irkutsk city.

The additional equipment such as TWSTFT mobile and stationary and Leica TDA 5005 is used too.

The two new generation laser stations were established in Mendeleevo and Irkutsk in 2018. Preliminary tests confirmed declared few millimeters accuracy (Fig. 10). This technique allows to provide time and frequency transfer experiments based on different technologies.

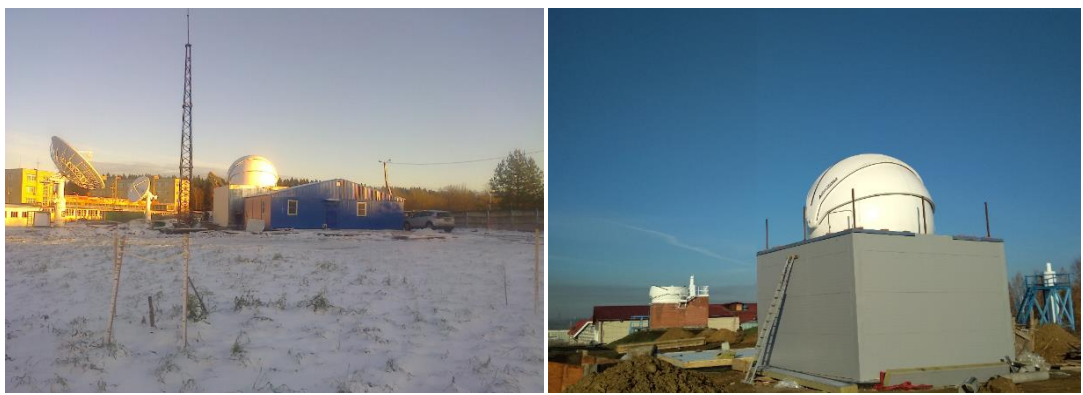

Fig. 10. SLR in VNIIFTRI and its East-Siberian branch toward to millimeters.

# Gravity Field

**Kaftan V.[1], Popadiev V.[2], Sermiagin R.[2], Sugaipova L.[3]**

[1]Geophysical Center of the Russian Academy of Sciences, Moscow, Russia
[2]Federal Scientific-Technical Center of Geodesy, Cartography and Spatial Data Infrastructure
[3]Moscow State University of Geodesy and Cartography, Moscow, Russia

Series of articles are devoted to the issue of processing satellite gradiometry data after the successful GOCE project [Нейман, Сугаипова, 2015a, 2015b, 2016a, 2017]. An algorithm of handling gradiometry data to receive harmonic coefficients of geopotential is proposed. The gradiometry data best fit for recovery of mid-frequency part of gravity signal. So, it is recommended to fulfil a bandpass filtering firstly. Approximation allowing downward continuation of measured second derivatives of the geopotential from real orbits to the mean-orbit sphere is implemented. The obtained information is transformed from a chaotic grid onto regular one on the same sphere. The ordered two-dimensional output array is convenient for the subsequent harmonic analysis. Possibility of application of classical sampling theorems of Gauss-Legendre (GL), Driscoll-Healy (DH) and relatively new theorems - sampling theorem of McEwen-Wiaux (MW) and the optimal sampling (OS) scheme theorem on the sphere for implementation of the harmonic analysis of a gravitational field of Earth is considered. Numerical experiments have shown that application of OS method is preferable as in comparison to other methods it requires the least information and has the same precision characteristics. All of these methods have shown sufficient accuracy of geopotential recovery and verified other authors recommendations on using grids 3-4 times finer than Nyquist frequency.

The short review of the satellite projects planned by the European and American space agencies and the research centers aimed at further specification of a gravitational field of Earth at different scales is given in [Сугаипова, 2015].

The monograph [Нейман, Сугаипова, 2016b] contains the key information on the multi-scale analysis and synthesis of signals in relation to gravitational field of the Earth. The most commonly used scaling functions, wavelets and radial basis functions are described. Their spatial localization allows to model high-frequency components of the geopotential effectively. For the harmonic analysis of a mid-frequency part of the gravity field possibilities of the beforementioned sampling theorems on the sphere and the corresponding schemes of numerical integration are summarised. The monograph also includes recommendations for joint processing of heterogeneous measurements by the least squares method. Theoretical considerations are illustrated by examples from geodetic practice.

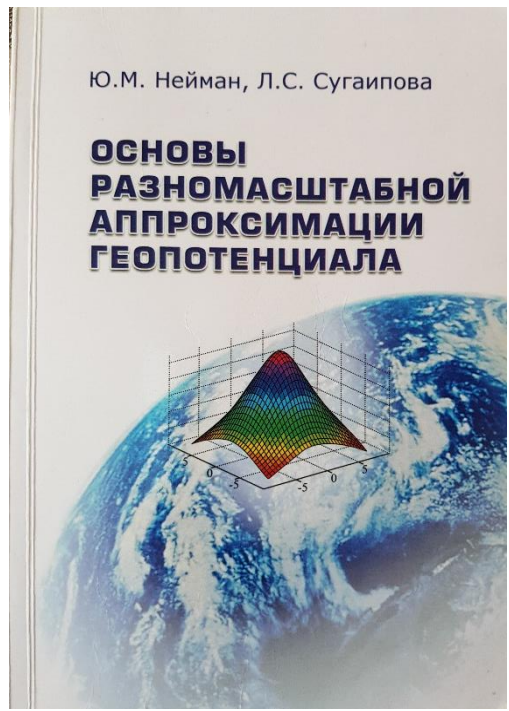

**Fig.** Monograph "Basics of multi-scale approximation of geopotential".

A brief overview on scaling functions and wavelets are provided in [Сугаипова, 2017a]. Expressions for gravity disturbance and gravity anomaly of the scaling functions and Abel-Poisson wavelets are derived. The algorithm of selection of the most effective poles of the spherical radial basis functions (SRBF) on the basis of geometrical interpretation of orthogonal least-squares method is described. The numerical experiment demonstrates application of the received formulas.

The problem of determining quasigeoid from gravimetry has a well-known solution by means of Stokes' integral. To solve the same problem using regional gravity data, different modifications of Stokes' kernel can be used. One of them was proposed by O.M. Ostach. In [Сугаипова, 2017b] *the frequency characteristic of the truncation operator of Stokes' kernel onto inner zone of $\psi_o$ radius* is introduced. There it is established that expansion of the truncated (by Ostach) Stokes' kernel into Legendre polynomials differs from the expansion of Stokes' kernel by this frequency characteristic. In [Сугаипова, 2018a] the fundamental equation of physical geodesy in local area is proposed to solve by means of spherical radial basis functions (SRBF) instead of integrating. New scaling functions and wavelets are introduced. These functions use the frequency characteristic of the truncation operator of Stokes' kernel that intercepts contribution of far zone. The two-step numerical experiment carried out to test these new scaling functions has shown a high precision of recovering height anomalies from regional gravity data.

Errors arising due to linearization and spherical approximation of the fundamental equation of physical geodesy can be weaken by replacement of the standard normal field generated by an equipotential ellipsoid. A low-frequency part of one of the modern global models of the geopotential can be used as a new normal field. In [Сугаипова, 2018b] it is described the iterative procedure of refinement of the standard normal field. The first 70 harmonics of the model GO_CONS_GCF2_DIR_R5 were used as a new normal field. Statistics of the corresponding numerical experiment have shown that the values of the disturbing potential and height anomaly decrease more than by 28 times when using the new normal field. Therefore, proposed refinement of a normal field can be of use when solving problems dealing with assumptions about smallness of disturbing potential, closeness of artificial telluroid to the real Earth etc. At the same time, additional computing expenses are insignificant. The iterative procedure converges in two-three steps and do not take much computational capability and time.

In the article [Попадьев, 2018], using the example of a strictly horizontal tunnel in a mountain massif, it is shown that normal heights have a fundamental advantage over orthometric ones, even if the latter are precisely known. The orthometric height along the level surface of the tunnel floor varies within 2.5 cm, while the normal height is almost constant. By projecting a measured potential difference to any part of a segment of a normal field line of force, one can obtain a set of types of normal heights, of which the normal orthometric (orthonormal) height is most known; and only the normal height according to Molodensky does not require additional knowledge of the geodetic height at any point. The question of choosing a system of heights is one of the main ones when establishing a general terrestrial system of heights; for a general terrestrial system of geopotential numbers, the development of a method for calculating anomaly in height is still relevant. Of interest is the following order: the calculation of the approximate value of the quasigeoid height from the solution of the classical Molodensky problem for mixed anomalies (the Brovar method of 1963), the conversion of mixed anomalies into pure anomalies, and then the solution of the "GPS problem" for pure anomalies by the 1971 Brovar anomaly (last it is possible to perform several times, since the magnitude of the anomalous potential must be the same). This approach allows full use of the potential difference contained in the mixed anomalies, and knowledge of the edge surface contained in the pure anomalies. Calculations on models with conical mountain ranges show the absolute accuracy of calculating the elements of the anomalous field of 3 cm using mixed anomalies and 6 cm using pure anomalies.

The small value of the normal heights correction allows also to use now the global gravity field models for its computations, for long (several hundred kilometers) leveling lines, the total correction, an integral value, will be weakly sensitive to errors of the individual point anomalies [Попадьев, Кулиев, 2017]. The total correction to the line Krasnovodsk — Charjew 1165 km long was —14.8 mm for free-air anomalies restored from the EGM-2008 Bouguer anomaly ($N = 2160$), c.f. the catalog value —14.2 mm. The total correction to the leveling line Tikhoretsk — Beslan and Beslan — Khasav-Yurt 685.6 km long (North Caucasus) is 62.6 mm, when the value from the catalog is 63.4 mm.

With further increase in the accuracy of navigation using onboard inertial navigation systems, it is of interest to correct the readings of onboard accelerometers taking into account the full gravity vector calculated over the earth's surface at flight altitude. So far, it is sufficient to introduce a correction using the grid of ground-based anomalies in free air and the existing schemes for evading the plumb line [Попадьёв и др., 2015].

International comparisons of absolute gravimeters were carried out to assess their metrological characteristics at Russian (Pulkovo, Svetloye, Lovozero, Zvenigorod, TsNIIGAiK) and Finnish (Metsähovi) reference gravimetric sites [Mäkinen et al., 2015; 2016a; 2016b]. Gravimeters FG5 (manufactured in the USA, Micro-g Lacoste), GBLP 001, GBL-M 002, GABL-PM, GABL-M (made in Russia, IAE SB RAS) participated in the comparisons. The new ballistic gravimeter GABL-M was tested especially in the Arctic conditions. The comparisons were purposefully carried out at sites (Fig. 10) with different physiographic conditions [Mäkinen et al., 2016a; 2016b]. Metrological estimates of the accuracy of all types of gravimeters were obtained and the levels of their accuracy and the possibility of using them in precision measurements were confirmed.

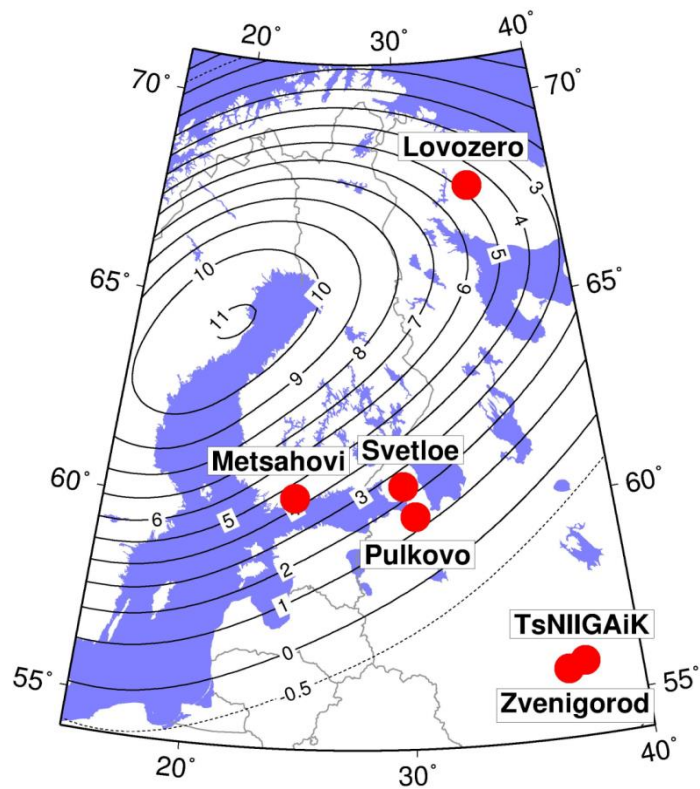

Fig.10. Absolute gravity stations of the Russian-Finnish geodetic study

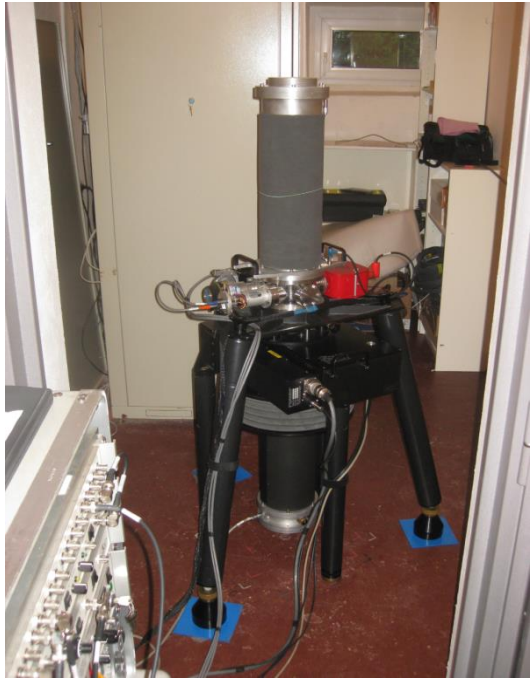

FG5-110 (Russia)

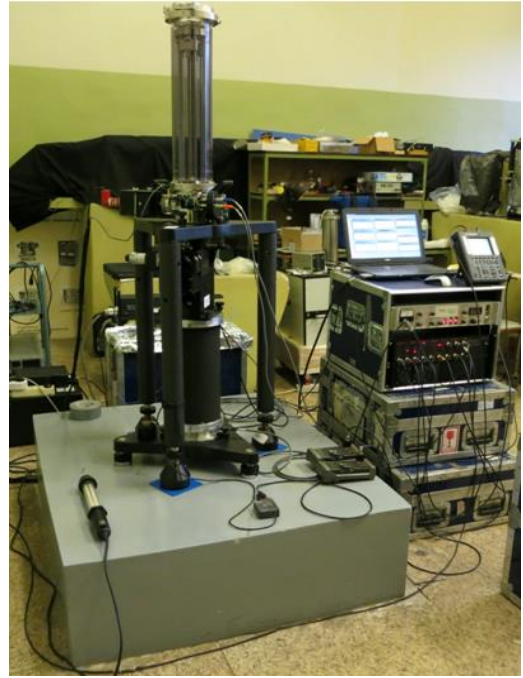

FG5X-221 (Finland)

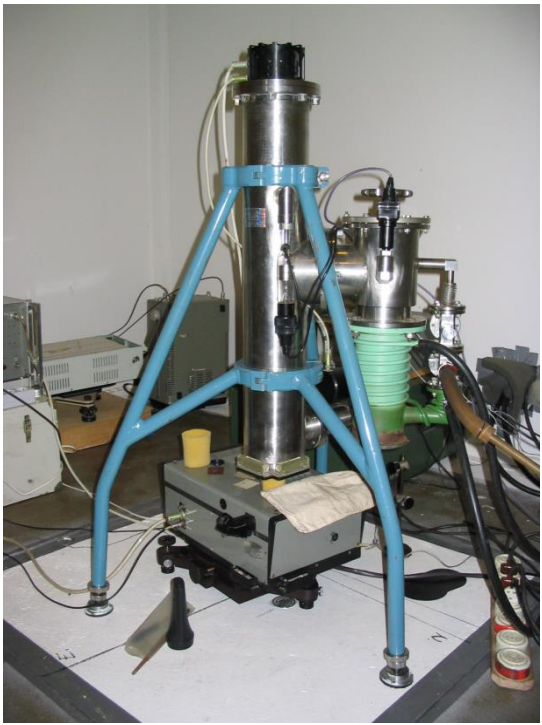

GBLP-001 (Russia)

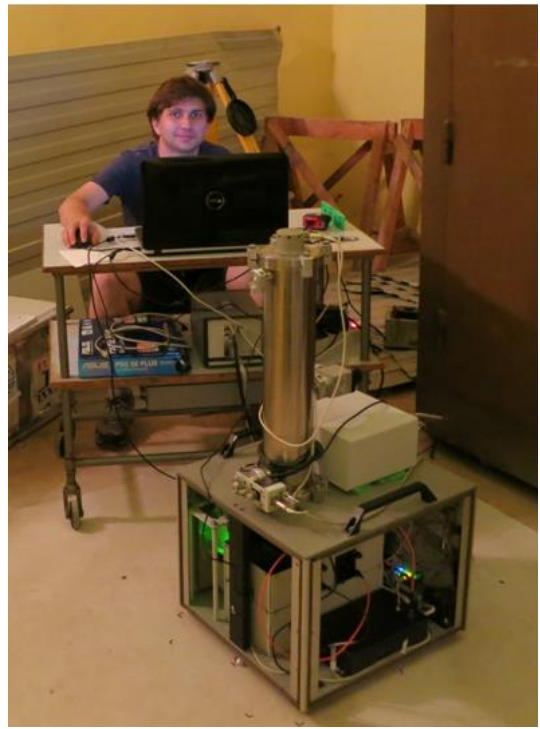

GBLM-002 (Russia)

Fig.12. Absolute gravity meters of Finland and Russia.

# Geodynamics

Gerasimenko M.[1,4], Gorshkov V.[2], Kaftan V.[3], Malkin Z.[2], Mazurov B.[4], Shestakov N.[1,5], Steblov G.[6]

[1]Institute of Applied Mathematics, FEB RAS, Vladivostok, Russia
[2]Pulkovo Observatory, Saint Petersburg, Russia
[3]Geophysical Center of the Russian Academy of Sciences, Moscow, Russia
[4] Siberian State University of Geosystems and Technologies, Novosibirsk, Russia
[5]Far Eastern Federal University, Vladivostok, Russia
[6]Geophysical Survey of the Russian Academy of Sciences, Obninsk, Russia

Based on the statistical analysis of spatiotemporal distribution of earthquake epicenters and perennial geodetic observation series, new evidence is obtained for the existence of slow strain waves in the Earth. The results of investigation allow to identify the dynamics of seismicity along the northern boundary of the Amurian plate as a wave process. Migration of epicenters of weak earthquakes ($2 \leq M \leq 4$) is initiated by the east-west propagation of a strain wave front at an average velocity of 1000 km/yr. It was founded a synchronous quasi-periodic variation of seismicity in equally spaced clusters with spatial periods of 3.5 and 7.26 degrees comparable with the length of slow strain waves. The geodetic observations at GPS sites in proximity to local active faults show that in a number of cases, the GPS site coordinate seasonal variations exhibit a significant phase shift, whereas the time series of these GPS sites differ significantly from a sinusoid. Based on experimental observation data and the developed model of crustal block movement, it was founded that there is one possible interpretation for this fact that the trajectory of GPS station position disturbance is induced by migration of crustal deformation in the form of slow waves [Bykov, Trofimenko, 2016].

An analysis of the spatial-temporal evolution of the deformation of the earth's surface in the San Francisco Bay area from continuous GPS measurements over more than 10 years revealed the migration of full shear deformation along the Hayward and Calaveras fault systems. Kinematic visualizations of the deformation process show the accumulation of significant deformations in the region of the epicenters of a pair of moderate earthquakes (M5.45 & M5.6) on October 31, 2007 in the Southeast part of the Hayward fault zone two years after the events. The displacement of the deformation front with a magnitude of ~ 3 * 10-6 ($\sigma$ = 10-7) from the place of its origin to the place of the large Napa earthquake (M6.02) on August 24, 2014 proceeded at a speed of ~ 20 km / yr. A visual analysis (see Fig. 1) suggests that the cumulative deformation of ~ 3 * 10-6, which has accumulated and reached the place of the future strong shock, was the trigger of the large Napa earthquake (August 24, 2014). The result obtained demonstrates a high probability of the existence of a new phenomenon — the migration of the deformation of accumulated earthquake-induced stress [Кафтан, Мельников, 2018b].

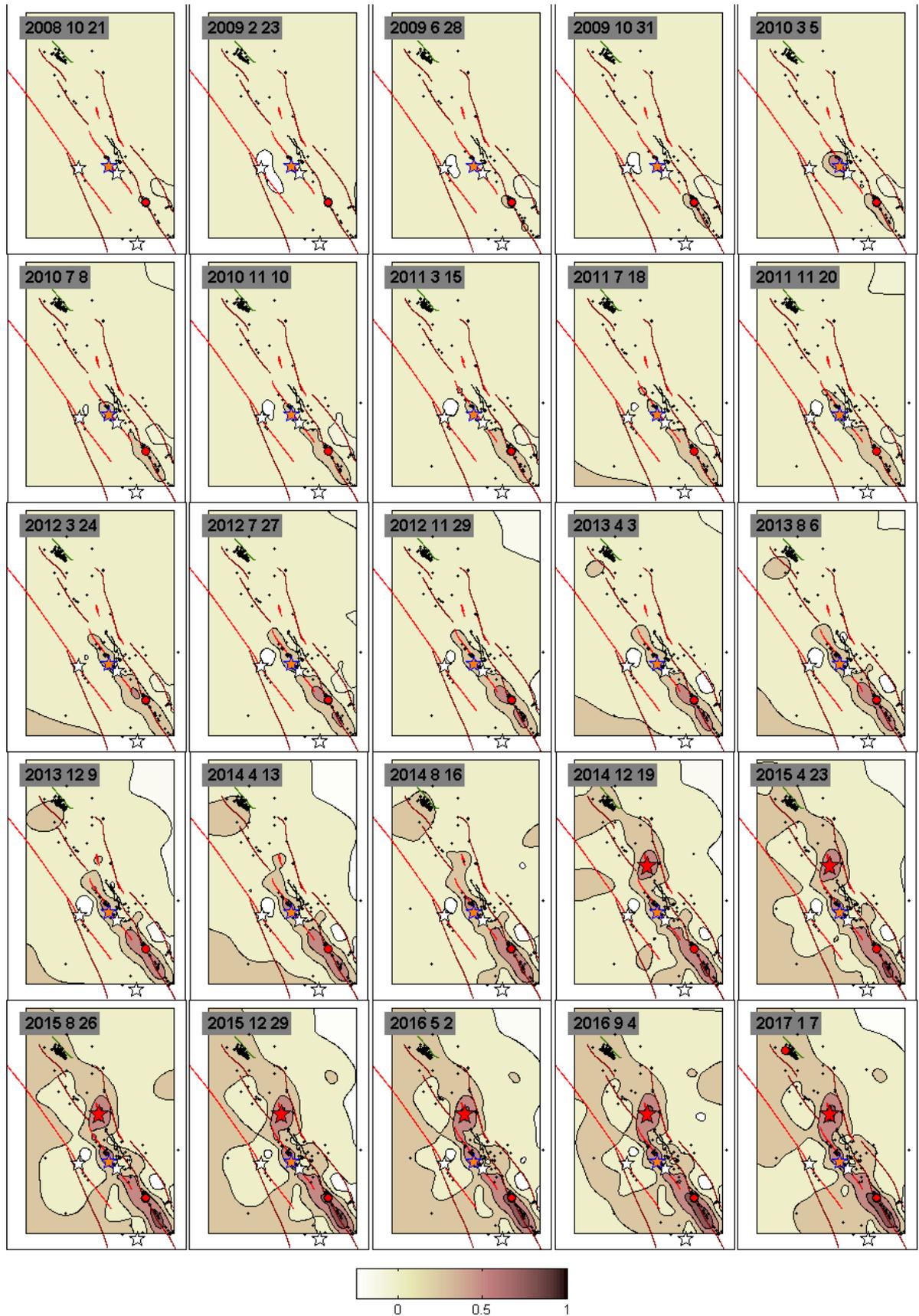

Fig. 1. Spatio-temporal distribution of full shear deformation. Isolines are drawn through $0.25 * 10^{-5}$. The big red star is the Napa earthquake M = 6.0. Small white stars - the strongest historical earthquakes. A small orange star is an expected catastrophic earthquake [4, 5]. Red circles - moderate earthquakes M = 5.4-5.6. Black dots - small earthquakes 3 <M <5. Fault zones sown by red, brown, and green lines.

The new insights into the nature of seasonal variations in coordinate time series of GPS sites located near active faults and methods of their modeling are provides. Monthly averaged coordinate time series were analyzed for several pairs of collocated GPS sites situated near the active fault intersection area, in close proximity to the central part of the northern boundary of the Amurian plate and the vicinity of the San Andreas Fault zone. It is concluded that the observed seasonal variations are best described by a breather function which is one of the solutions of the well-known sine-Gordon equation. The obtained results suggest that, in this case, the source of seasonal variations may be caused by the appearance of solitary strain waves in the fault intersection system, which may be qualitatively treated as standing waves of compression-extension of the geological medium. Based on statistical testing, the limits of applicability of the suggested model have been established [Trofimenko et al., 2016].

The nature of long-term vertical crustal movements observed in the southern Primorye is oscillatory and related with the subduction zone processes [Герасименко и др., 2016]. Vertical postseismic displacement velocities ranging from 7 to 14 mm/yr have recognized up to January 2014 at a set of GPS stations located ~1000 km westward away from the Tohoku earthquake epicenter. They have been continuing. Vertical coseismic crustal offsets were not detected in the far-field zone. Change of the sign of the earth's surface tilt in the far field zone might be used as a long-term precursor of catastrophic earthquakes in the subduction zone.

Parameters of the strain rate tensor were calculated for the territory of the Upper Amur region from the published GPS observations. As a result, such parameters as principal strain rates and principal directions, maximum shear strain rates and its directions, the dilatation rate, the second invariant of the strain rate tensor, have been obtained. The results indicate a high tectonic activity in the interaction zone of the Eurasian, Amurian lithospheric plates and the Stanovoy geoblock. Areas of high-speed aseismic displacements are revealed. The development of the Eurasian and Amurian plate boundaries at the present stage suggests that NE-striking faults will be activated [Ашурков и др., 2018].

In 2006 the geodynamic GNSS monitoring network was developed on the Kuril Islands [Prytkov et al., 2017a, 2018b]. Today the network consists of 11 stations for continuous and regular recording, covering the island arc quite evenly. Fig. 2 shows the rates of the geodynamic network stations in relation to the North American plate. In the central part of the island arc, post-seismic shifts of the Earth's surface have been observed since the double Simushir earthquakes in 2006 Mw = 8.3 and in 2007 Mw = 8.1. They are oriented towards the deepwater trench. The fading transient process caused by viscoelastic stress relaxation in the asthenosphere and the upper mantle has continued for the past ten years.

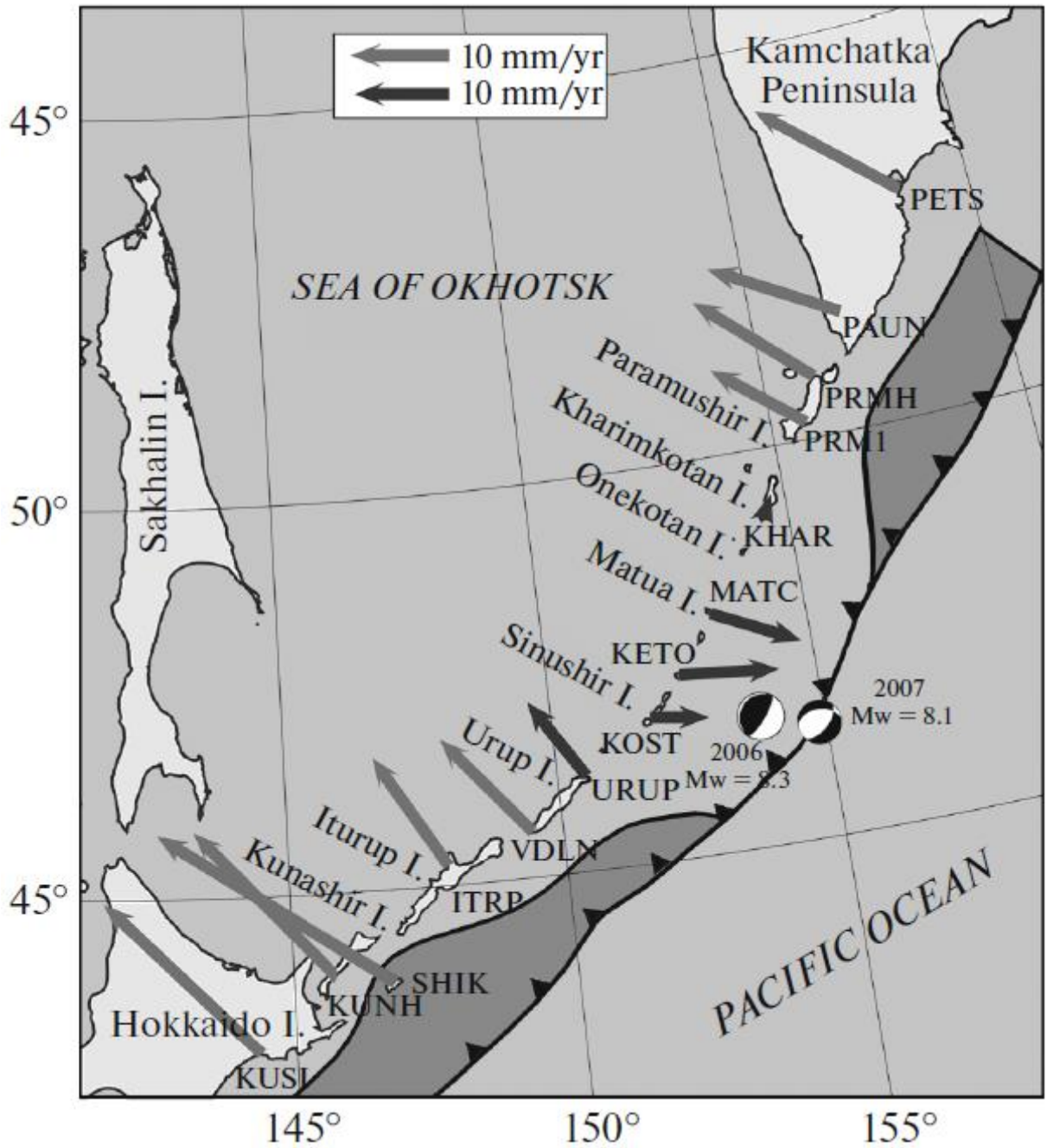

Fig. 2. Movement rates of GPS/GLONASS stations of the Kuril geodynamic network and neighboring districts. Light arrows show interseismic rates of stations of the Kuril geodynamic network for the period of 2007-2015. Dark arrows show station movements between June 2014 and June 2015. The root mean-square errors of rate determinations do not exceed 2 mm. The solution of the geometry of lithosphere plate coupling is provided in the form of the width of the inclined contact zone approximated by the spline function. In the central part of the island arc the mechanisms of the sources of the Simushir earthquakes are shown.

The earth surface deformation was modeled for the North, Central and South Sakhalin on the basis of deformation velocities recorded by the GPS stations of the Sakhalin Geodynamical Network. A pattern of contemporary horizontal deformation is intricate in the vicinity of the main submeridional faults of the island. On the island surface, the dominant deformation regime is compression; however, the spatial distribution of deformation is heterogeneous. The horizontal compression is mainly sublatitudinal and SW-NE trending. In addition to compression, there are zones of rather intense right-lateral strike-slip in the northern and central parts of the island, while stretching dominates in the south-eastern parts. The regional geodynamic setting is reflected in the seismicity of the island. Recently, the seismic activity has been increased in the areas characterized by intensive surface deformation, while the areas of low deformation rates correlate with the zones of weak and sparse seismicity [Прытков, Василенко, 2018].

On August 14, 2016 at 11: 15 am UTC near the western coast of the central Sakhalin Island at a depth of ~9 km an earthquake of magnitude Mw = 5.8 occurred. The intensity of impacts in the epicenter zone reached 7 on the MSK-64 scale. Based on the total seismotectonic data, it was found that the movement in the seismic focus occurred along the transverse fault connecting two large regional fault zones of the submeridional stretch – Western Sakhalin and Central Sakhalin. The seismic moment tensor is determined by the inversion of body waves and the earthquake focus modeling is performed. Subhorizontal compressive stresses of the NE-SW direction have been found in the seismic focus, which are consistent with the recent deformation behavior of the central part of the island. The focal mechanism type is a reverse fault of the southwest direction with a minor shear component. The reconstructed displacements in the earthquake focus made it possible to calculate the coseismic deformations of the earth's surface of the epicenter zone [Прытков и др., 2018].

The results of statistical simulation for the seismicity within the zone of interaction among the Amur, Eurasian, and Sea-of-Okhotsk plates where our seismological analysis revealed the Olekma-Stanovoi seismic area, the Tan Lu zone, and the northern Sakhalin-Japan island arc are presented. We discuss the spatial irregularities in the seismicity maxima distribution in annual cycles. It was found that when considered in clusters that are ordered over longitude, the seismicity maxima form long chains of anomalies. It is shown that the mean migration rate of seismicity maxima does not vary throughout the east-west trending Olekma-Stanovoi and the Tukuringra-Dzhagdy seismic zones that travel from the north-south eastern framing of the Amur plate as far as the east-west Tukuringra-Dzhagdy seismic zone [Trofimenko et al., 2015a].

Three recurrent spatial cycles have been observed. The clusters with similar distribution of earthquakes are suggested to alternate being equally spaced at 7.26 degrees (360-420 km). A comparison of investigation results on the structure of seismicity in various segments of the Amurian microplate reveals the identity between the alternation pattern observed for meridional zones of large earthquakes and a distinguished spatial period. The displacement vector for seismicity in the annual cycles is determined, and the correspondence between its E-W direction and the displacement of the fronts of large earthquakes is established. The elaborated model of seismic and deformation processes is considered, in which subsequent activation of clusters of weak earthquakes ($2 \leq M \leq 4$), tending to extend from the Japanese-Sakhalin island arc to the eastern closure of the Baikal rift zone, is initiated by the displacement of the strain wave front [Trofimenko et al., 2017b].

Seismic anisotropy has been studied using data of S and ScS waves from earthquakes occurred in the mantle transition zone and recorded at stations located on the Asian continental margin, the Island of Sakhalin and the southern part of the Kamchatka Peninsula. Mantle seismic anisotropy beneath the Amur Plate is studied using the data of ScS waves reflected from the outer core of the Earth from local deep earthquakes in the area of five stations located in Primorye and Priamurye regions. The results of measuring the parameters of the split ScS waves near the stations show the dominance of the polarization azimuths of the fast wave along the eastern directions and are consistent with azimuthal anisotropy and the direction of motion of the Pacific Plate (300 degrees) and Amur Plate (similar to 120 degrees) depending on the epicenter-station direction. In southern Kamchatka, the azimuths of fast S and ScS waves from the M 8.4 Okhotsk earthquake have been determined along the Pacific plate motion. The azimuths of fast S waves from the aftershocks of the Okhotsk earthquake and the 2008-2009 large events are sub-parallel to the trench strike. In Primorye, in the vicinity of the TEY station, the polarization azimuths of the fast ScS wave dominate in the interval of NE-E directions orthogonally to the lines of the mantle flow along the complex 3D surface of the subsiding Pacific Plate. It is revealed that the delay time of ScS waves increases to 2 s and 3.4 s in the upper mantle and in the mantle transition zone, respectively, as the depth of events increases [Luneva, Pupatenko, 2016a, 2018b].

Well-pronounced TEC disturbances with an average period of about 10 min and amplitude of 0.07–0.5 TECU (total electron content unit, 1 TECU = 10 16 el m À2) were detected [Perevalova et. al., 2015]. These disturbances were initiated by an ionospheric source activated about 5–6 min after the airburst. The disturbance velocity damping with time and distance from the airburst was revealed using a GPS interferometric technique. Several modes of TEC disturbances with propagation velocities ranging from 250 to 660 m/s were distinguished through the distance-time diagram analysis. The estimated position of ionospheric source of TEC disturbances is shifted 36 km southwestward from the airburst, which is most likely to be associated with the conjoint influence of TEC data errors, damping and anisotropy of TID propagation velocity.

It is shown [Berngardt et. al., 2015] that 6–14 min after the bolide explosion, GPS network observed the cone-shaped wave front of traveling ionospheric disturbances (TIDs) that is interpreted as a ballistic acoustic wave. The typical

TIDs propagation velocity were observed 661 ± 256 m/s, which corresponds to the expected acoustic wave speed for 240 km height. Fourteen minutes after the bolide explosion, at distances of 200 km, we observed the emergence and propagation of a TID with annular wave front that is interpreted as gravitational mode of internal atmospheric waves. The propagation velocity of this TID was 337 ± 89 m/s which corresponds to the propagation velocity of these waves in similar situations. At EKB radar, we observed TIDs in the sector of azimuthal angles close to the perpendicular to the meteorite trajectory. The observed TID velocity (400 m/s) and azimuthal properties correlate well with the model of ballistic wave propagating at 120–140 km altitude. It is shown that the azimuthal distribution of the amplitude of vertical seismic oscillations with periods 3–60 s can be described qualitatively by the model of vertical strike-slip rupture, propagating at 1 km/s along the meteorite fall trajectory to distance of about 40 km. These parameters correspond to the direction and velocity of propagation of the ballistic wave peak by the ground. It is shown that the model of ballistic wave caused by supersonic motion and burning of the meteorite in the upper atmosphere can satisfactorily explain the various azimuthal ionospheric effects, observed by the coherent decameter radar EKB, GPS receivers network, and the azimuthal characteristics of seismic waves at large distances.

Results of the study of far-field propagation of the main types of traveling ionospheric disturbances (TIDs) caused by the March 11, 2011, Japan earthquake are presented [Перевалова и др., 2016]. The study relies on double frequency phase data acquired by several GPS networks operating in the Asia-Pacific region: the FEB RAS complex geodynamic network, the GS RAS Kuril network, the Southern Korean network, the international IGS network http://sopac.ucsd.edu), and the GEONET networks in Japan (http://mekira.gsi.go.jp). It was determined the direction and distance of propagation for various TID types and have revealed TID attenuation asymmetry in different directions from the epicenter: the TIDs are registered up to 2000–2200 km from the epicenter northwestward, but they fade at a distance of 1300 km northeastward. The TIDs damp rapidly in the azimuth sector of 10–45º from the epicenter (at ~800 km). Fast TIDs (V~2000–3000 m/s) effectively propagate southwestward from the epicenter; they are less pronounced northwestward and are absent northeastward. The TIDs caused by acoustic waves (V~600–1000 m/s) are propagating symmetrically in all directions, but almost disappear at 800–1500 km from the epicenter. Only this type of TIDs is observed northeastward. Slow TIDs (V~150–300 m/s) move largely southwestward and northwestward.

Coseismic gravity changes stem from (1) vertical deformation of layer boundaries with density contrast (i.e., surface and Moho) and (2) density changes of rocks at depth. Tanaka et. al. [2015] reported that they have been observed in earthquakes with Mw exceeding ~8.5 by Gravity Recovery and Climate Experiment (GRACE) satellites, but those of M8 class earthquakes have never been detected clearly. Coseismic gravity change of the 24 May 2013 Okhotsk deep earthquake (Mw8.3) smaller than the detection threshold. In shallow thrust faulting, factor (2) is dominant, while factor (1) remains secondary due to poor spatial resolution of GRACE. In the 2013 Okhotsk earthquake, however, factor (2) is insignificant because they occur at depth exceeding 600 km. On the other hand, factor (1) becomes dominant because the centers of uplift and subsidence are well separated and GRACE can resolve them. This enables GRACE to map vertical ground movements of deep earthquakes over both land and ocean.

Zhao et al., [2018] re-estimated the seismic moment of the 2011 Mw 9.0 Tohoku earthquake by using far-field co-seismic displacements from 41 continuous GPS stations in Northeast Asia. In support of nearly 100 GPS campaign stations, they interpolate precise inter-seismic velocities at continuous GPS stations which are lacked of pre-seismic observations by using the Gaussian distance-based interpolation method, aiming to pursuing more accurate post-seismic displacements. Then, they inverted the mantle viscosity and effective lithosphere thickness in the study area by employing non-linear least squares fitting of the high signal-to-noise horizontal post-seismic displacements from 2.5 to 4.5 years after the earthquake at 9 GPS stations distributed perpendicular to the fault plane of the earthquake, and the viscoelastic spherical Earth dislocation theory with modified fault slip model. The optimum solution of the inversion realizes the viscosity being $1.0 \times 10^{19}$ Pa·s and the lithospheric thickness being 30 km. The far-field post-seismic displacements caused by afterslip and mantle viscoelastic relaxation are approximately equal within ～2.5 years, and after 4.5 years, the cumulative effects of the latter component dominate the total far-field post-displacements. The result indicated that we cannot ignore the the effects of mantle viscoelastic relaxation in the far field, even immediately after this earthquake. In addition, in the study of the far-field post-seismic displacements, we should take the signal-to-noise ratio (SNR) of the GPS data into account with adequate observations to ensure the accuracy of the inversion.

The results of noise modeling in high-rate GNSS time series are presented by [Пупатенко, Шестаков, 2017]. GNSS data processing was performed using Precise Point Positioning (PPP) approach in real time mode. Initial data for noise modeling includes 10–13 days of one's observations from 128 global distributed continuous GNSS stations belong to IGS and CORS networks. Modified sidereal filtering was applied to coordinate time series for reducing multipath error. The LRTGNM (Low Real-Time GNSS Noise Models), MRTGNM (Median Real-Time GNSS Noise Models) and HRT-GNM (High Real-Time GNSS Noise Models) was developed based on statistical distribution of GNSS time series noise. The models was compared with Peterson's seismic noise models NHNM (New High Noise Model) и NLNM (New Low Noise Model) as well as with high-quality accelerometers noise models AHNM (accelerometric high noise model) and ALNM (accelerometric low noise model). It is shown that for periods 300 s and less the both seismic instruments are more sensitive than the GNSS receiver.

The analysis of the data obtained during the last 30 years has revealed a number of features that indicate the modulation of high-frequency seismic noise (HFSN) by tides and connection of the HFSN with various geophysical processes, including changes in the stress state of the medium during the preparation of earthquakes [Saltykov et al., 2018]. An important property of the tidal influence upon the HFSN was found: the HFSN response is not stable over time. In the 1990s, based on the results received in Kamchatka, a hypothesis about the connection between the variations of the tidal component of the HFSN and the geodynamic processes in the region was proposed. Later, based on long-term field observations, it was shown that the tidal sensitivity of the HFSN is most stable and statistically significant during the preparation of large local earthquakes.

The investigation results of the April 12, 2014, M = 4.5 Primorye earthquake (Far Eastern Russia) Primorye is a region with quite moderate shallow seismicity which has been insufficiently investigated so far in this respect [Shestakov et al., 2018]. Based on the obtained instrumental data of regional seismic networks and macroseismic data collected in southwestern Primorye on the crustal earthquake with M = 4.5 occurred on April 12, 2014, it was first succeeded in determining the hypocenter parameters and the focal mechanism of the mainshock of this shallow earthquake and estimating the hypocenter parameters of the following aftershock.

Theoretical infinite time viscoelastic response represents spatial postseismic displacement features, such as; 1. Near -field: coseismic > postseismic; 2. Middle - field (with distance equal to fault with): coseismic = postseismic; 3. Far-field coseismic < postseismic. This indicates that subduction great earthquake might cause significant impact to far field continental deformation field. Relaxation time constant of postseismic displacement is controlled by viscosity. Estimated viscosity of $10^{19}$ Pa·s predicts 200-400 yrs time constant. This duration is approximately equal to recurrence interval of M>8.5 giant subduction earthquakes in Kamchatka, Kuril and Japan trenches. Analytical interseismic and postseismic displacement in continent indicate the 1cm/yr westward interseismic velocity and more than 50cm postseismic displacement by the 2011 Tohoku earthquake, respectively. These facts suggest that past great subduction earthquakes, e.g. 1952 Kamchatka M9.0, 17th century Kuril M > 8.8, and pre-historical events recorded in tsunami deposits, also might generate regional transient postseimic deformation with 200-400yrs time constant. Reconstruction of deformation time history might be available if we have proper viscoelastic parameters and subduction earthquake history [Takahashi et al., 2018].

For the calculation of the viscoelastic relaxation, subsurface rheological structure modeling is necessary because it controls the pattern of the postseismic displacement field and its duration. It was assumed a simple two-layered spherical earth model, which consists of upper elastic and lower Maxwell viscoelastic layers. Based on the information of the seismic velocity structure and postseismic deformation signal of the 2011 Tohoku earthquake observed at the IGS sites in northeast Asia, it was estimated appropriate rheological structure (65 km of elastic thickness and 1.5×10^19 Pa·s of viscosity of the viscoelastic layer).Along the Japanese Island 15 of historical large earthquakes (M ≥ 8.0) have been recorded since 869A.D. In order to check the effect of their postseismic deformation signals, it was calculated surface displacement field for ~1300 years with the estimated rheological structure. As a result, most of earthquakes may contaminate the stable displacement field around the coastal side of the Asian continent with the movement velocity of several mm/year after each event. The displacement signal from each postseismic deformation depends on the magnitude of earthquake, focal distance, fault mechanism, duration, and so on. Thus, their time series of total displacement shows complex pattern at each area [Ohzono et al., 2018].

A set of continuous and semi-continuous GNSS sites belonging to FEB RAS, FEFU and private companies were used for monitoring of postseismic movements. The maximum cumulative value of postseismic displacements has already exceeded 90 mm which is approximately twice grate than the appropriate coseismic shifts. The most intense nonlinear movements occurred during the first 6 months after the mainshock. Since that time, postseismic displacement rates were stabilized and can be fitted well by the liner regression. The orientation of postseismic velocity vectors varies from the south-east for the southern part of the region to the pure south for the northern group of GNSS stations. Relatively simple model with Maxwell's rheology quite well explaining observed the far-field late postseismic displacements [Shestakov et. al., 2018].

Analysis of geological and geophysical data on the boundary zones between the Eurasian plate and other plates shows poor knowledge of the deep structure of the region. A comprehensive analysis of the deep seismic sounding (DSS) materials and seismic data shows a significant expression of this boundary zone both in the deep Earth's crust structures and in the Moho. A zone of anomalous seismicity and deep structure extends along the DSS profile for several hundred kilometers. It was refined the position of the main boundary between the Eurasian and Okhotsk plates, which passes approximately along 144° E [Solov'ev et al., 2016].

Current relation between atmospheric pressure and vertical displacements was estimated at the center of Siberian Anti Cyclone with size varied from 2000 km to 3000 km. Pressure-displacement coefficients (PDC) can be achieved by three years observation (0.997 mm/mbar for NVSK GPS station). It is used for elastic module study of geology medium with maximum thickness up to 600 km. In the context of elastic model, the modulus of rigidity is estimated to be 113 GPa. Annual vertical changes were obtained by leveling near the dam of the reservoir. PDC ratio was 1.15 mm/bar for these places. In elastic theory, the Young modulus was calculated by sixteen years of leveling measurements. This result can effectively be represented for upper crust. The results were checked by solution for coseismic displacement of Chyia-Altai earthquake (Sep. 27, 2003,Mw7.3). Coseismic results calculated by static modules agree with experimental coseismic GPS data at 10% level [Timofeev et al., 2017a].

Gravity observation is interpreted together with GPS observation data which was obtained from 2012 to 2015 at the same station. The goal of the observation was the investigation of gravity variation with time and seismicity situation monitoring. Gravity observation was developed at special basement by absolute gravimeter (GABL type) and by spring gravimeter (SCINREX CG-5and gPhone type). Tidal models were tested by results of observation with spring gravimeters. Reduction task was solved, as the experimental data received from different points of Shults Cape Observatory was used. Applied reduction coefficient is 203.3 mGal m1, and agrees with theoretical calculation. Gravity anomaly varied from 30 mGal to 46 mGal, which also depend on difference reference system. Experimental results were used for testing of the structure of continental boundary, which also depends on the sea bottom flexion. Thickness of elastic layer was estimated from12 km to 18 km by using different models [Timofeev et al., 2017, 2018a, 2018b].

Significant changes, e.g., in the coseismic displacements of the Earth's surface are recorded in the zones of large earthquakes. These changes should manifest themselves in the variations of gravity. The effects of the catastrophic мw = 9.0 Tohoku, Japan, earthquake of March 11, 2011 were identified in Primorye in the far zone of the event. The empirical data are consistent with the results of modeling based on the seismological data. The coseismic variations in gravity are caused by the combined effect of the changes in the elevation of the observation point and crustal deformation [Timofeev et al., 2017a, 2018a].

The Aleutian and Kuril-Kamchatka arcs meet at a triple junction of the Pacific (PAC), Bering (BER), and North American (NAM) plates. It was inverted GPS observations from the westernmost Aleutian (Komandorsky) Islands and Kamchatka for the fault locking depth and block motion in the far western Aleutian transform boundary. Three boundary models were considered: (1) only the Aleutian thrust fault without a trench-normal component, (2) only a strike-slip fault in the back arc north of the Komandorsky Islands, and (3) a rigid Komandorsky sliver bounded by the Aleutian and back-arc faults. Observed velocities prefer Model 3, with a secular westward sliver velocity of 51 mm/a relative to NAM (two thirds of the total PAC-NAM motion). The observed velocities are ~10% slower because of elastic strain from boundary faults. The best fitting locking depth of faults bounding the sliver is 12 km, which is similar to depths observed in diverse tectonic environments [Kogan et al., 2017].

The most important practical significance of studying of geodynamic processes is the solution of problems of the forecast, reduce risk, and reduce the effects of geodynamic catastrophes of natural and technogenic of the impact, environmental monitoring. Multidisciplinary inverse problem in the multidimensional complex measurements to find the properties of the medium at the specified information about the fields. Solving the inverse problem as multidisciplinary, comprehensive geodetic and geophysical measurements, we can obtain new qualitative results. An example of the inverse problem is also the need to distinguish between actual vertical displacement of the earth's surface points and the displacement of the water level surface, horizontal displacement of these points and change the direction of the plumb line in time. Мазуров [2017] show the solution to the inverse problem in a volcanic field from geodetic measurements, but with the increase intralesional pressure in the accumulation of magma in the upper magma chamber of the volcano.

Levin et al., [2017] analyzed the relationship between variations of the Earth's rotation rate and the geodynamic processes within the Earth's body, including seismic activity. The rotation rate of a planet determines its uniaxial compression along the axis of rotation and the areas of various surface elements of the body. The Earth's ellipticity variations, caused naturally by the rotation rate variations, are manifested in vertical components of precise GPS measurements. Comparative analysis of these variations is considered in view of modern theoretical ideas concerning the Earth's figure. The results justify further research that is of interest for improvement of space systems and technologies.

A destructive and strong earthquake of magnitude Mw 7.2 occurred in the eastern part of Turkey that known as Van region on October 23, 2011. In the study [Kaftan, Kaftan, Gök, 2018] the earth's surface deformation characteristics before, during and after the Van earthquake were analyzed. The data from permanent GPS network with an area 300x300 km2 were used to evaluate deformations of the earth's surface around the epicentral region of the earthquake in the time interval January 16, 2009 - October 29, 2012 with a daily temporal resolution. Dilatation deformations, total shear, horizontal and vertical displacements were determined approximately 3 years before and one year after the 2011 Van earthquake. To observe the change of the deformation process before and after the earthquake, determined results for each day were compiled as a movie. Unusual deformations were discovered several months before the earthquake at a distance of about a hundred kilometers from the future epicenter. The aim of this study is to identify deformation earthquake precursors from GPS data with a daily resolution and wide coverage of the study area. Kinematic features of deformation of the earth's surface in connection with the Van earthquake, seismic regime and tectonic structure of the region were obtained.

Attempts are being made to identify the deformation precursors of strong earthquakes according to continuous GNSS observations in regions of high seismic activity. For this purpose, daily coordinate or double difference GPS solutions are used. For each day, the values of deformations of total shear and dilatation are obtained from the values of horizontal displacements of geodetic points. The time series of horizontal displacements occurring the years before strong seismic events are used. Deformations are calculated within the finite elements - triangles of geodetic networks, formed according to the Delunae rule. Due to the spatial heterogeneity of the distribution of geodetic points, the dimensions of the finite elements vary greatly. This leads to incorrect interpretation of the results. To eliminate this effect, the strain values are scaled relative to the average area of the triangles of the control network. Before strong earthquakes of the Northwest coast of North America and Eastern Turkey, in the vicinity of their epicenters, kinematic visualizations of the deformation process were obtained. This made it possible to trace the evolution of deformations from the beginning of observations to a seismic event and after. Attempts to identify the growth of deformations before the seismic events in the immediate vicinity of future epicenters were not crowned with success. Anomalous inhomogeneities of deformation in fracture zones, probably involved in the seismic process, which precede the main shocks are revealed. These anomalies exceed the background characteristics of deformations several times. They are confined to fault lines and epicenters of weak earthquakes in the area of observation, which suggests their connection with the preparation of the main seismic event [Кафтан, Мельников, 2016; 2017; 2018а; Kaftan, Melnikov, 2016a; 2016b; 2016c; 2017; 2018; Kaftan, Kaftan, Gök, 2018].

The results of the study of horizontal deformations and vertical displacements in the area of the active volcano of Etna (Sicily Island, Italy), obtained from observational data of global satellite navigation systems in 2011–2017 with the registration interval one day using rarely located stations of the regional geodetic network presents in publications [Kaftan, Rodkin, 2018; Kaftan, Rodkin, 2019].

Studies of Etna are especially important for the following reasons.

Localization of a volcano in an area with a high population density.

Almost continuous eruption.

Discrepancy of the location of the volcano plate-tectonic hypotheses.

Subregional tendencies of deformation of the studied territory are revealed. Extension strains were recorded not only in the summit crater area, but also at a considerable distance from it in the aquatory of the Ionian Sea. The latter suggests the existence of an extensive deep-sea supply system of the volcano with sources far distanced from the summit active crater. The results of geological and geophysical studies of the coastal and underwater areas of the region are discussed. The feasibility of studying the deformations according to observational networks is demonstrated, albeit with a low density of stations, but with a large coverage area.

The results of the study of recent movements and deformations of the earth's crust in the area of future disposal of radioactive waste are presented in [Kamenev et al, 2018; Татаринов и др., 2016а; 2016b; 2018; Tatarinov et al., 2017].

Some studies devoted to theoretical and technological aspects of geodetic geodynamics.

The adjustment questions of repeated geodetic measurements performed for the study of crustal deformation or engineering structures was examined [Герасименко, Каморный, 2014]. Itis shown that in the separate adjustment systematic errors will not affect the determination of the displacement vector only under the same weights in both periods of measurement, but the mean square error of unit weight will be distorted. This disadvantage is eliminated in the adjustment of differences of measurements. What is shown theoretically the reliability assessment, contrary to popular belief, is not deteriorated.

Provides an example of how to use the thematic mapping when studying geodynamic processes in terms of forecasting natural and industry-related disasters [Панжин и др., 2015, Мазуров и др., 2016]. Visualization of displacement of the Earth's surface points based on the results of the series of geodetic and gravimetric measurements allows more found to allocate active geological structures, blocks, tectonic faults. This knowledge is necessary to predict the locations of possible seismic events and take appropriate preventive measures to ensure the security of the population, industrial objects, etc. Evidence-based assessment of stressedstrained state of the array in the area of influence of the quarries of Kachkanar mining. An identified grouping vectors movements in clusters, which largely coincide with the tectonic structures. Fixed measurements in three cycles of alternating nature of the trend and cyclical deformations allows to draw conclusions about the presence in array migration concentration zones of deformation.

The scientific significance of the researches of geodynamic processes is to obtain new knowledge about the Earth, its structure, evolution, various physical fields (gravitational, magnetic, etc.), the spatial and temporal structure of the physical surface. The most important practical significance of studying of geodynamic processes is the solution of problems of the forecast, reduce risk, and reduce the effects of geodynamic catastrophes of natural and technogenic of the impact, environmental monitoring. Multidisciplinary inverse problem in the multidimensional complex measurements to find the properties of the medium at the specified information about the fields. Solving the inverse problem as a multidisciplinary, comprehensive geodetic and geophysical measurements, we can obtain new qualitative results. An example of the inverse problem is also the need to distinguish between actual vertical displacement of the earth's surface points and the displacement of the water level surface, horizontal displacement of these points and change the direction of the plumb line in time. The article [Мазуров, 2015b, 2017a, 2019] shows the solution to the inverse problem in a volcanic field from geodetic measurements, but with the increase intralesional pressure in the accumulation of magma in the upper magma chamber of the volcano.

Geodetic data and their subsequent statistical analysis make it possible mathematical modeling and identification of the stress-deformed state of geodynamic systems in respect of the aspect of prediction of natural and man-made disasters. Geodetic monitoring of geodynamic processes necessary for solving a number of scientific and practical tasks of geodesy - expanding and maintaining the national geodetic network, studying changes in gravity field in time, using GNSS technology. Most important continue ahead of research is mathematical modelling of geodynamic systems in a predictive order. For the study of complex (nonlinear) geodynamic processes should be selected corresponding to mathematical framework. Here are some of the theoretical foundations of the study of rotation movements of the earth's surface. Mentioned mathematical model rotary circular structures of the Earth. There [Мазуров, 2016b, 2019] are mathematical models, accounts that are contributing to the sudden nature of global, regional and some local geodynamic processes. They are based on differences in temporal and spatial scales, of geodynamic systems. Theoretical bases of the description of rotational motions on a plane by a system of differential equations. When examples of integral curves which can be qualitative characteristics of geodynamic systems. In many cases, a similar trajectory corresponds to the rotational horizontal movements of the earth's surface.

Geodetic measurements provide important and statistically estimated information about the coordinates of geodetic points and their changes over time. This geodesic information can be used to study geodynamic processes and their manifestations, primarily on the earth's surface. Particularly intense such geodynamic phenomena occur in areas of active development of minerals due to the intense man-made effects on the near-surface layer of the Earth. It is logical to perform the description of surface motions using mathematical field theory. According to the changes of geodetic elements (coordinates, heights, directions) after repeated measurements it is possible to imagine the field of displacement vector of geodetic points. When studying the stress-strain state of the earth's surface, the vectors obtained can be used not only to calculate the earth's deformation tensor in the area under study, but also the differential characteristics of the vector field. One of them is called divergence. The authors of the article [Колмогоров и др., 2018] propose to determine the divergence of the vector field of surface displacements by discrete geodesic observations of displacement vectors made only on the surface of the study area. The model of the vector displacement field can be chosen taking into account the set of source data and the density of placement of geodetic points, the carriers of spatial coordinates.

The possibilities of local geodynamic monitoring using GNSS (GPS, GLONASS) were investigated [Kaftan, Sidorov, Ustinov, 2015; Kaftan, Sidorov, Ustinov, 2015]. Close accuracy of both systems is demonstrated. The accuracy is expected to increase approximately one and half times with an increase of the number of GLONASS satellites.

Technological methods for improving the accuracy of local geodynamic monitoring are considered in [Kaftan, Ustinov, 2015].

Paradoxes of the comparative analysis of ground-based and satellite geodetic measurements in recent geodynamics are discussed in the study [Kuzmin, 2017]. Problems of the study of active tectonic faults using modern geodetic measurements are described and discussed in [Kuzmin, 2015; Кузьмин, Фаттахов, 2018; Фаттахов, 2017].

The velocity vector field of the dense regional GNSS network provides the basic data for the analysis of deformation processes in the faults, for study of the intraplate structure of region and other geodynamical studies. For research of these problems the database of GNSS station velocities (VDB) approximately covering the East European Craton (EEC) was created in 2015 (Gorshkov at al., 2015a) and supported now with colleagues of various geodetic companies (Gorshkov at al., 2017b, 2018a). At the beginning of 2019 the VDB has uniformly processed data for more than 360 GNSS stations (fig. 3, 4).

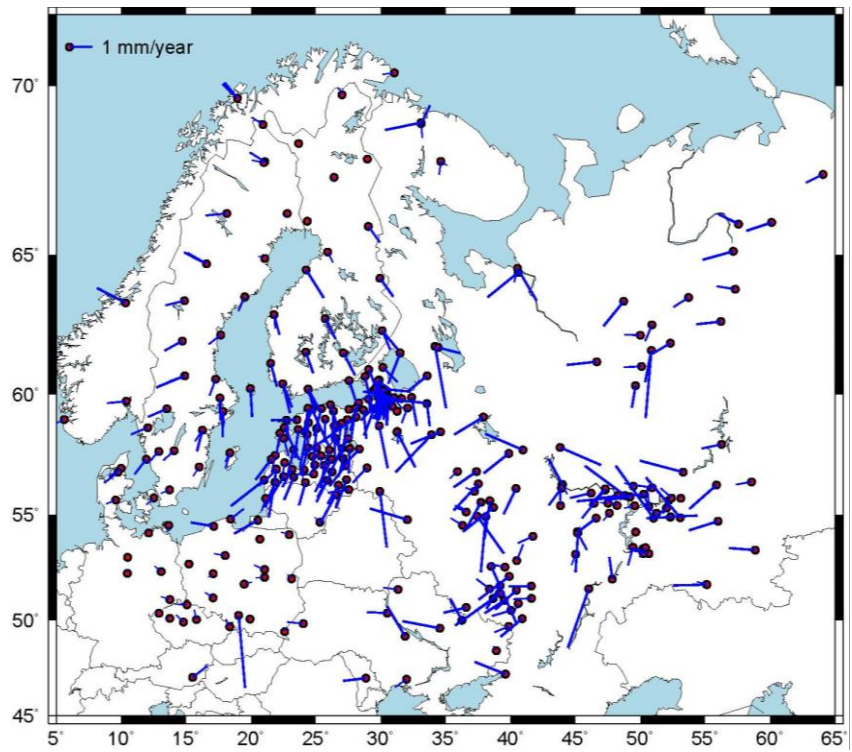

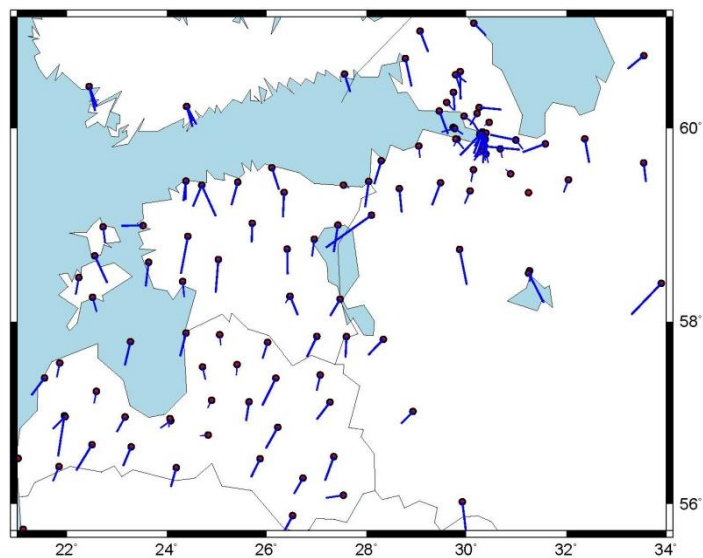

Fig.3. The horizontal velocity of VDB after removing Eurasian plate rotation according to the model (Altamimi, Z., L. Metivier, X. Collilieux. ITRF2014 plate motion model // Geophys. J. Int. (2017) 209, 1906–1912)

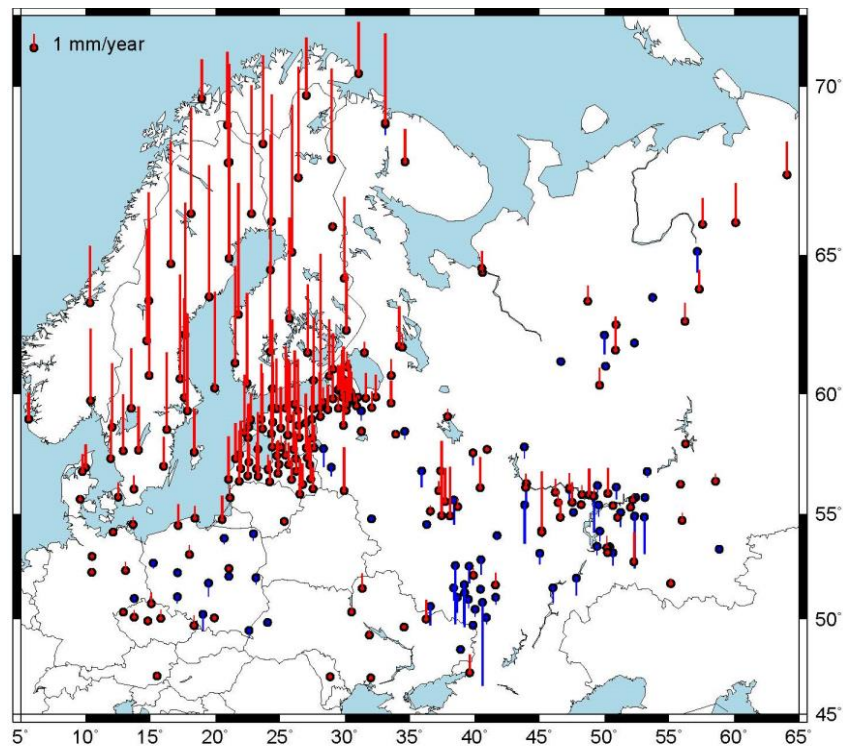

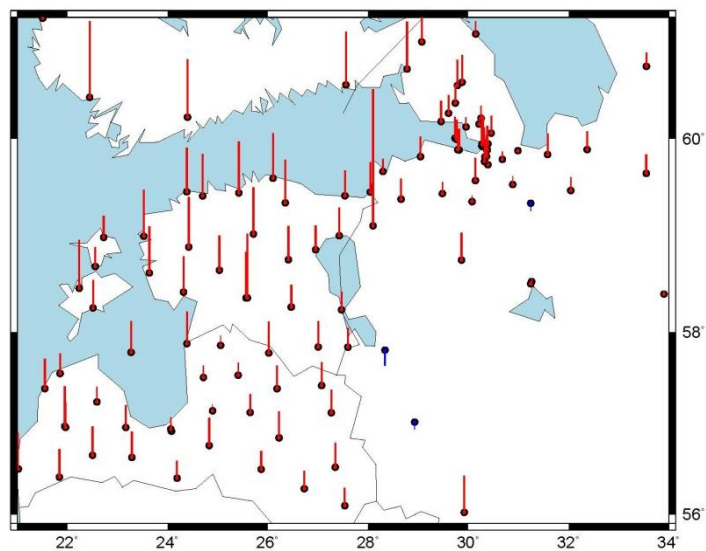

Fig.4. The vertical velocity of VDB.

The VDB is available on the website of the Pulkovo observatory (http://www.gaoran.ru/russian/database/station/databasev_eng.html), where the full description of used technique is also provided.

This VDB was used to develop research methods (Mokhnatkin at al., 2015) and for some geodynamical researches. So, the velocity of all VDB stations corrected for the postglacial rebounding by model (Peltier, W.R. Global Glacial Isostasy and the Surface of the Ice-Age Earth: The ICE-5G (VM2) Model and GRACE // Ann. Rev. Earth and Planet. Sci. 2004. 32. P. 111-149) was used for the assessment of the plate rotation (Gorshkov at al., 2015; Mokhnatkin at al., 2016; Gorshkov at al., 2018b). This assessment closely matches with global one by (Altamimi, Z., L. Metivier, X. Collilieux. ITRF2014 plate motion model // Geophys. J. Int. (2017) 209, 1906–1912). No significant mutual rotation was detected between Baltic shield and East-European platform when used current (2019) state of the VDB. However the general structural feature of the EEC velocity field was revealed, which is characterized by compression of the central areas of the EEC from NE and NW directions.

The accuracy of velocity GNSS stations is a most significant parameter for correct geodynamical conclusions. The estimation of informal velocity errors was made by using all compactly located GNSS-stations within small area (less 1 km base) in international GNSS nets (Gorshkov, Scherbakova, 2015b). The upper limit of these errors for more than hundred GNSS-stations was assessed by means of comparison of these station velocities. Approximately half of stations located in the same place have a significant velocity discrepancy up to 1mm/year in horizontal components and up to 3mm/year in vertical one. Therefore, the probability of such errors in any GNSS stations is equal 50%.

The analysis of the GNSS position series of the regional networks indicates that the position variations have often simultaneously in the whole region (hundreds km) impulse-shaped in-phase disturbance (so-called common-mode error, CME), which have up to 20 mm amplitudes and can last up to several weeks. These errors can distort the velocity field in regional scale and therefore have to be considered in geodynamical researches. The influence of various meteorological parameters (the atmospheric pressure and temperature, thickness of snow layer) on the appearance of these CMEs was searched by the GNSS and meteorological data in Baltic region of the VDB (Gorshkov, Scherbakova, 2017a). These CMEs often correlate with at atmospheric pressure and other meteorological parameters (fig. 5). That is evidence of imperfection of applied loading models. The CMEs appear more often in wintertime and partly could be explain by accumulated snow level, which is not used in the loading models. A number of the loading models have been compared in (Gorshkov, Scherbakova, 2016).

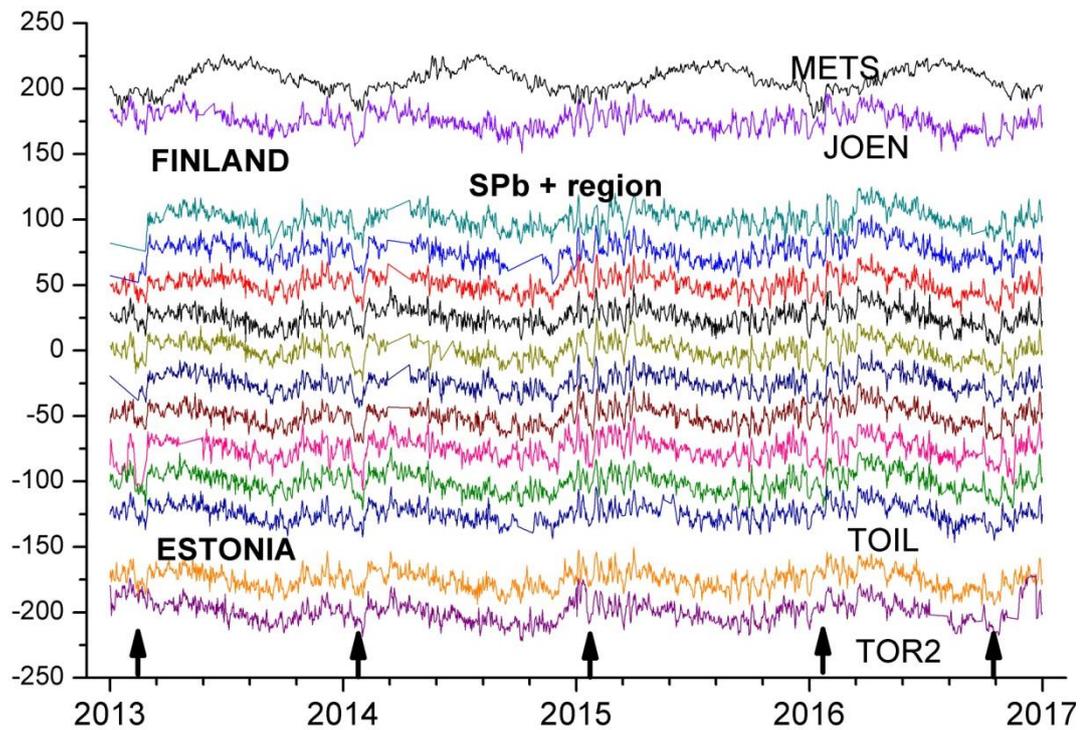

Fig. 5. Vertical (in mm) components of some GNSS stations. Arrows mark CME events

The analysis of long-term geophysical processes was made in regional scale by using current GNSS observations and newly reprocessed old (up to 2010) latitude observations in Kazan region (Mubarakshina at al., 2018) and Poltava one (Khalyavina L.Ya., Zalivadny, 2018). The correlation between latitude anomalies and seismic conditions was revealed in both region. In Poltava was also revealed decadal variations of the local vertical and the 12-year non-polar cyclicity, which correlates (r=0.7) with the solar activity index (fig. 6).

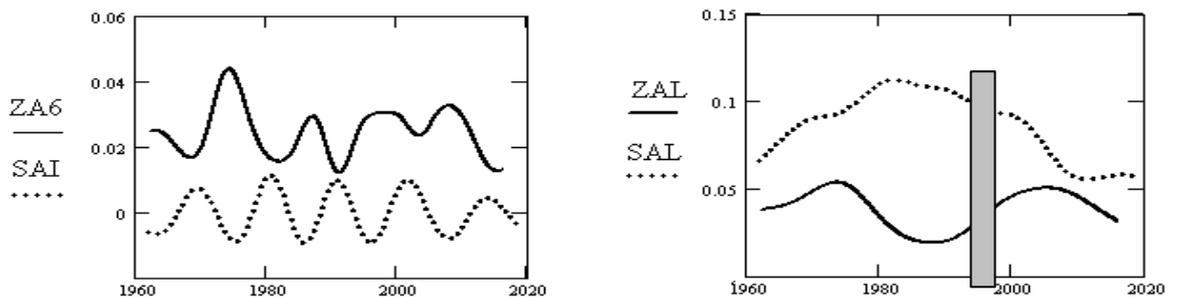

Fig. 6. The comparison of vertical variation (ZA6, ZAL) from non-polar variation of latitude with solar Wolf's number (SAI, SAL) in various band of spectrum (left – right).

The geophysical by-product of GNSS observation were presented in (Vorotkov at al., 2018), where the dynamics of water vapor content was studied in the troposphere over the Baltic-Ladoga region, and in (Trofimov at al., 2018), where the parameters of the ionosphere by GNSS-observations at Antarctic station Vostok were evaluated.

The impact of the seasonal signals in the station displacement on the celestial reference frame and Earth orientation parameters has been studied [Krasna et al, 2015]. The estimate has been made of empirical harmonic models for selected stations within a global solution of all suitable VLBI sessions. Mean annual models have been created by stacking yearly time series of station positions, which are then entered a priori in the analysis of VLBI observations. The results reveal that there is no systematic propagation of the seasonal signal into the orientation of celestial reference frame but position changes occur for radio sources observed non-evenly over the year. On the other hand, the omitted seasonal harmonic signal in horizontal station coordinates propagates directly into the Earth rotation parameters causing differences of several tens of microarcseconds.

It was earlier found that the variations in free core nutation (FCN) are connected with various processes in the Earth's fluid core and core-mantle coupling, which are also largely responsible for the geomagnetic field variations, particularly the geomagnetic jerks (GMJs). A new study of this effect revealed a new evidence of this connection [Malkin, 2016]. The large FCN amplitude and phase disturbance occurred at the epoch close to the newly revealed GMJ 2011. This event occurred to be the second large change in the FCN amplitude and phase after the 1999 disturbance that is also associated with the GMJ 1999. Moreover, the long-time FCN phase drift had changed suddenly in 1998--1999, immediately before the GMJ 1999, and seemed to change again at the epoch immediately preceding the GMJ 2011. The FCN amplitude showed a general long-time decrease before GMJ 1999, and it subsequently grew until GMJ 2011, and then seemed to decrease again. A smaller FCN change can be observed at the epoch around 2013, which is also suspected as the GMJ epoch. The latter confirms the suggestion that a rapid change in the FCN amplitude and/or phase can be used as an evidence of the GMJ that is not clearly detected from the geomagnetic observations.

The impact of the change in the International Celestial Reference System (ICRS) from the first ICRF (International Celestial Reference Frame) second extension ICRF-Ext.2 (ICRF1), to the last ICRF realization (ICRF2) has been investigated [Tornatore et al., 2015]. At this aim, authors have processed a set of VLBI experiments during 27 years to estimate session-wise station coordinates, their velocities, baseline lengths and Celestial Pole Offsets (CPO) components, using as radio source reference catalog once ICRF1 and once ICRF2. The authors have analyzed time series of estimated geodetic parameters and their formal errors, and the time series of the differences of the parameters estimated in the two reference source catalogs. For the analysis of the time series, they have used Allan Deviation (ADEV) and its modifications. The findings confirm that the switchover from ICRF1 to ICRF2 yields improvements for example in the baseline lengths repeatability and in the scatter of CPO series of about 2 uas. However, the results highlight also some discrepancies: the series of station coordinate differences (ICRF1-ICRF2) show significant noise at mm level and the series of baseline length differences has a bias of about 2 mm, then a time variable residual signal is present in the CPO difference time series. Figure 7 shows the improving in the VLBI baseline length repeatability obtained with ICRF2 w.r.t. ICRF1.

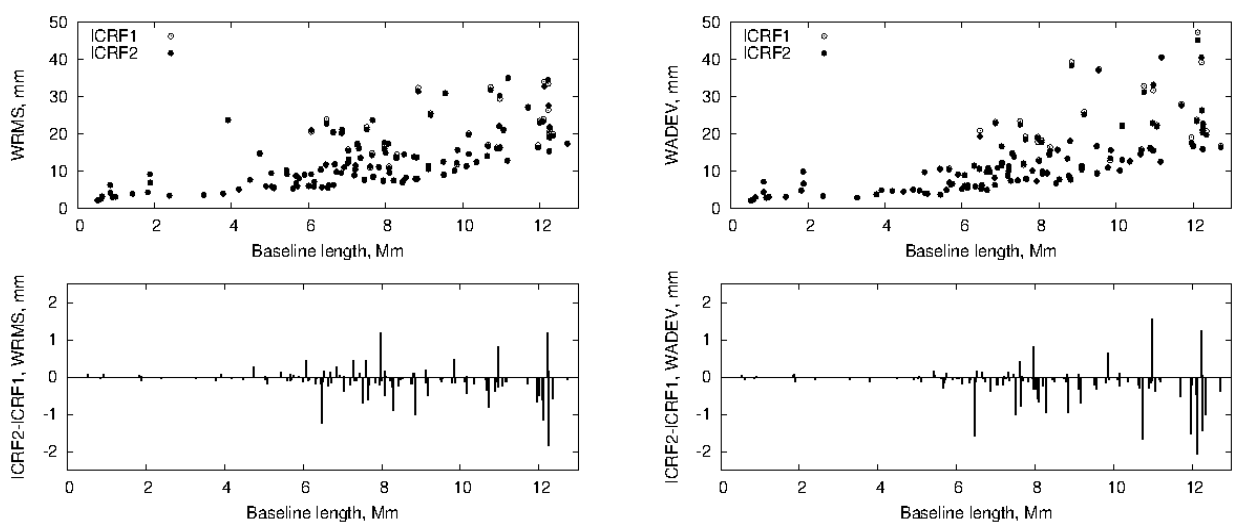

Fig. 7. Baselines length repeatability (top) and the difference between the baselines length repeatability estimates computed with ICRF2 and ICRF1.

The series of publication are devoted to the geodynamic studies at the oil-gas deposits and critical engineering objects [Kuzmin, 2019; Квятковская и др., 2017; Кузьмин, 2018; ].

Фаттахов Е. А. Спектрально-временной анализ светодальномерных наблюдений на Камчатском и Ашхабадском геодинамических полигонах, Вестник СГУГиТ, 2017, Т. 22, № 4, С. 5–17.

Berngardt O. I., Perevalova N. P., Dobrynina A. A., Kutelev K. A., Shestakov N.V., Bakhtiyarov V. F., Kusonsky O. A., Zagretdinov R. V., Zherebtsov G. A. (2015) Towards the azimuthal characteristics of ionospheric and seismic effects of "Chelyabinsk" meteorite fall according to the data from coherent radar, GPS and seismic networks, Journal Geophysical Research. Space Physics, Vol. 120, Iss.12, P. 10754-10771. DOI:10.1002/2015JA021549

Bondur V.G., I.A. Garagash, M.B. Gokhberg, The dynamics of the stress state in Southern California based on the geomechanical model and current seismicity: Short term Earthquake prediction, Russian Journal of Earth Sciences, Vol. 17, ES105,doi: 10.2205/2017ES000596, 2017.

Bondur V.G., M.N. Tsidilina, E.V. Gaponova, O.S. Voronova, Joint analysis of various precursors of seismic events using remote sensing data at the example of earthquake in Italy (24.08.2016, M6.2), 17-th International Multidisciplinary Scientific GeoConference SGEM 2017, 29 June – 5 July, 2017. Albena, Bulgaria,pp. 149 - 162.

Bondur V.G., O.S. Voronova, Using remote sensing data to monitor volcanic activity: Mount Etna case study, 17-th International Multidisciplinary Scientific GeoConference SGEM 2017, 29 June – 5 July, 2017. Albena, Bulgaria, pp. 339 - 346.

Bykov V.G., Trofimenko S.V. (2016) Slow strain waves in blocky geological media from GPS and seismological observations on the Amurian plate, Nonlinear Processes in Geophysics, 2016, V. 23, № 6, P. 467-475.

Gabsatarov Y., I. Vladimirova, L. Lobkovsky, I. Garagash, B. Baranov, G. Steblov, Analysis of the tectonic deformations in the Chilean subduction zone caused by the 2010 Maule earthquake on the basis of satellite geodetic data, Исследования по геоинформатике: труды Геофизического центра РАН, 2017, Т. 5, № 1,pp. 60, DOI: 10.2205/CODATA2017

Gorshkov V., V. Kaftan, Z. Malkin, N. Shestakov, G. Steblov, Geodynamics. In: Savinykh V.P., Kaftan V.I. (Eds.), National Report for the IAG of the IUGG 2011-2014, Geoinf. Res. Papers, 2015, vol. 3, BS3005, GCRAS Publ., Moscow, pp. 37-58, DOI: 10.2205/2015IUGG-RU-IAG

Gorshkov V., N. Scherbakova, S. Smirnov, A. Mohnatkin, S. Petrov, D. Trofimov, T. Guseva, N. Rosenberg, V. Perederin, Deformation of the South-Eastern Baltic Shield from GNSS observations, Proc. of International conference Journees-2014 "Recent developments and prospects in ground-based and space astrometry", Editors: Z. Malkin, N. Capitaine. SPb, 2015 - 262 p. ISBN 978-5-9651-0873-2,pp. 211- 214.

# Earth's Rotation

Gorshkov V.[1], Malkin Z.[1], Pasynok S.[2]

[1]Pulkovo Observatory, Saint Petersburg, Russia
[2]National Research Institute for Physical-Technical and Radio Engineering Measurements (VNIIFTRI), Mendeleevo, Moscow Reg., Russia

Joint analysis of celestial pole offset (CPO) and free core nutation time series was performed to investigate their actual accuracy and the mutual impact [Malkin, 2017]. Three combined celestial pole offset series computed at the Paris Observatory (C04), the United States Naval Observatory (USNO), and the International VLBI Service for Geodesy and Astrometry (IVS), as well as six free core nutation (FCN) models, were compared from different perspectives, such as stochastic and systematic differences and FCN amplitude and phase variations. The differences between the C04 and IVS CPO series were mostly stochastic, whereas a low-frequency bias at the level of several tens of microarcseconds was found between the C04 and USNO CPO series. The stochastic differences between the C04 and USNO series became considerably smaller when computed at the IVS epochs, which can indicate possible problems with the interpolation of the IVS data at the midnight epochs during the computation of the C04 and USNO series. The comparison of the FCN series showed that the series computed with similar window widths of 1.1 yr to 1.2 yr were close to one another at a level of 10 µas to 20 µas, whereas the differences between these series and the series computed with a larger window width of 4 yr and 7 yr reached 100 µas. The dependence of the FCN model on the underlying CPO series was investigated. The RMS differences between the FCN models derived from the C04, USNO, and IVS CPO series were at a level of approximately 15 µas, which was considerably smaller than the differences among the CPO series. The analysis of the differences between the IVS, C04, and USNO CPO series suggested that the IVS series would be preferable for both precession-nutation and FCN-related studies.

The accuracy of CPO predictions made in 2007–2017 was investigated in [Malkin, 2016a, 2018]. The accuracy of the actual celestial pole position predictions obtained from 2013 in U.S. Naval observatory (model USNO) and Pulkovo observatory (models ZM2 and ZM4). It was found that Pulkovo models provide substantially better prediction accuracy than USNO model. Model ZM4 performs better for the prediction length up to 7–8 months, whereas model ZM2 is preferable for longer predictions (Fig.1).

Evolution of the VLBI EOP precision and accuracy over the past 37 years was investigated [Malkin, 2016b]. More than 13 million observations (VLBI delays) has been obtained during the 37 years of geodetic and astrometric VLBI programs are stored in the International VLBI Service for Geodesy and Astrometry (IVS). Despite historical interest, analysis of statistics of these observations can be useful to solve practical tasks, such as, e.g., selection of an optimal data interval for various studies or investigation of EOP accuracy and precision on the network geometry and other factors.

Members of the Geodesy section actively participated in the activity of the IAU Commission 19 "Rotation of the Earth" [Malkin, 2015].

The tidal free Length-of-Day variations (LODs) and geophysical excitation of LOD by angular momentum of atmosphere (AAMf) and ocean (OAMf) are investigated in [Горшков,. 2015] in combination with solar activity (SA) indexes (SSN and F10.7). There were revealed

(i) seasonal and quasi-biennial oscillation of LODs are completely determined by AAMf and OAMf which are in turn determined by SA;

(ii) the main solar cycle (11 year) in LODs is not fully conditioned by AAMf and OAMf and hence should have additional source of excitation;

(iii) LODs and SA 5-6 years variations are correlative, negative before the middle of 1980 years and positive after this time when astronomical monitoring of the Earth rotation was replaced by space one;

(iv) decadal LODs variations are almost synchronous with secular geomagnetic ones – the deceleration of west drift of the geomagnetic field forestall on 4 ± 0.5 years the deceleration of the Earth rotation. The main period of both variations is approximately 65-70 years.

The EOP activities at VNIIFTRI can be grouped in three basic topics:
–providing GNSS and SLR observations at five metrological sites acting under the auspices of Federal Agency on Technical Regulating of Metrology (ROSSTANDART);
–processing GNSS, SLR, LLR and VLBI observation data for EOP evaluation and GLONASS satellites orbit/clock;
–combination of EOP series for evaluation of reference EOP values.

VNIIFTRI is provide GNSS and SLR observations at five metrological sites acting under the auspices of Federal Agency on Technical Regulating of Metrology (ROSSTANDART) for purpose of maintenance of terrestrial reference frames (ITRF, FAGS and others) and EOP service. The SLR observations of the Lageos-1,2 and Etalon-1,2 satellites are used for purposes of EOP evaluation. VNIIFTRI takes participation in GNSS and SLR observations of IGS and ILRS and provides observations for its own purposes.

The GNSS, SLR, and VLBI observation data are processed in daily mode in VNIIFTRI for EOP evaluation [Блинов и другие, 2018]. The results are formed in the EOP raws and SINEX files form.

The SLR processing software uses the neural net algorithm for prediction of the orbits of laser satellites [Цыба Е.Н., Н.А.Вострухов, 2018 a]. As result, VNIIFTRI SLR EOP achieved good accuracy (80 and 100 microseconds of arc for terrestrial pole coordinates) and small latency (only one day) [Цыба Е.Н., Н.А. Вострухов, 2018 b]. See also Fig. 1.

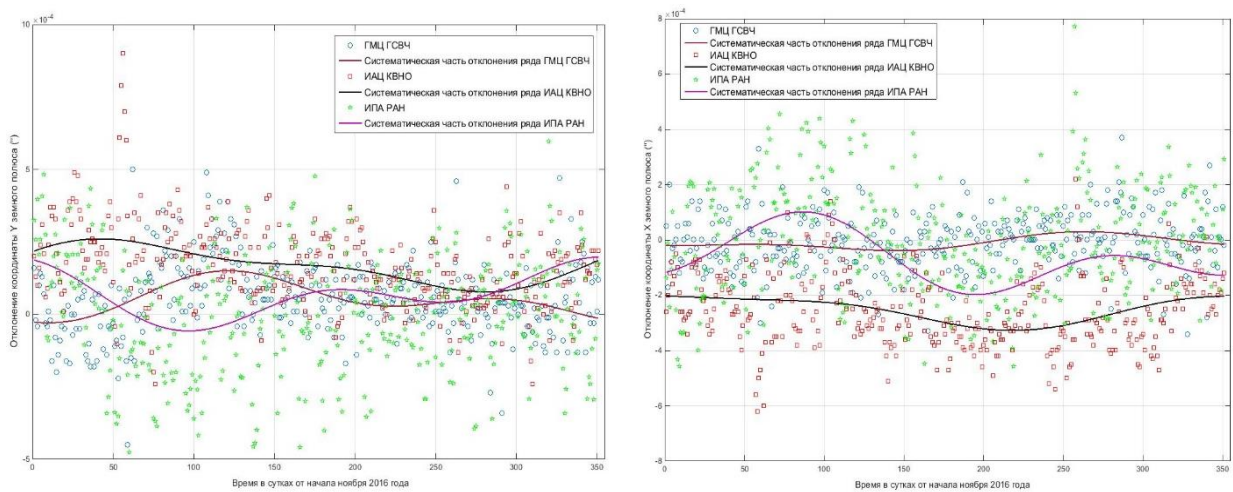

Fig. 1 – Differences of terrestrial pole coordinates VNIIFTRI (ГМЦ ГСВЧ, dark green circles for values and light brown line for smoothing values).

The modern program of UT1 evaluation based on Lunar Laser Ranging measurements was created in the MATLAB environment. Now only ILRS LLR data are processed, but it's ready for processing the Altay LLR station measurements too [Цыба Е.Н., Н.А.Вострухов, 2018 b].

For VLBI data processing the ARIADNA software package developed by V. Zharov (Sternberg Astronomical Institute of MSU, SAI) [V.E. Zharov, 2011]. Version 4.11 of this software was finished and tested at the end of 2018. All reductions are performed in agreement with the IERS Conventions(2010). Now package uses files in VGOS(NetCDF) and NGS format as input data and creates the SINEX output files for every IVS session. Package was automated for purposes Russian EOP operative service and now it is used in VNIIFTRI for operative VLBI sessions processing [Пасынок и др., 2019].

Results of UT1-UTC evaluation from VLBI data processing are shown on the fig. 2 and fig. 3.

The regular operative EOP evaluations using GPS data from Russian regional network (about 30 stations) are carried out using BERNESE 5.0 GNSS software and scripts developing in VNIIFTRI.

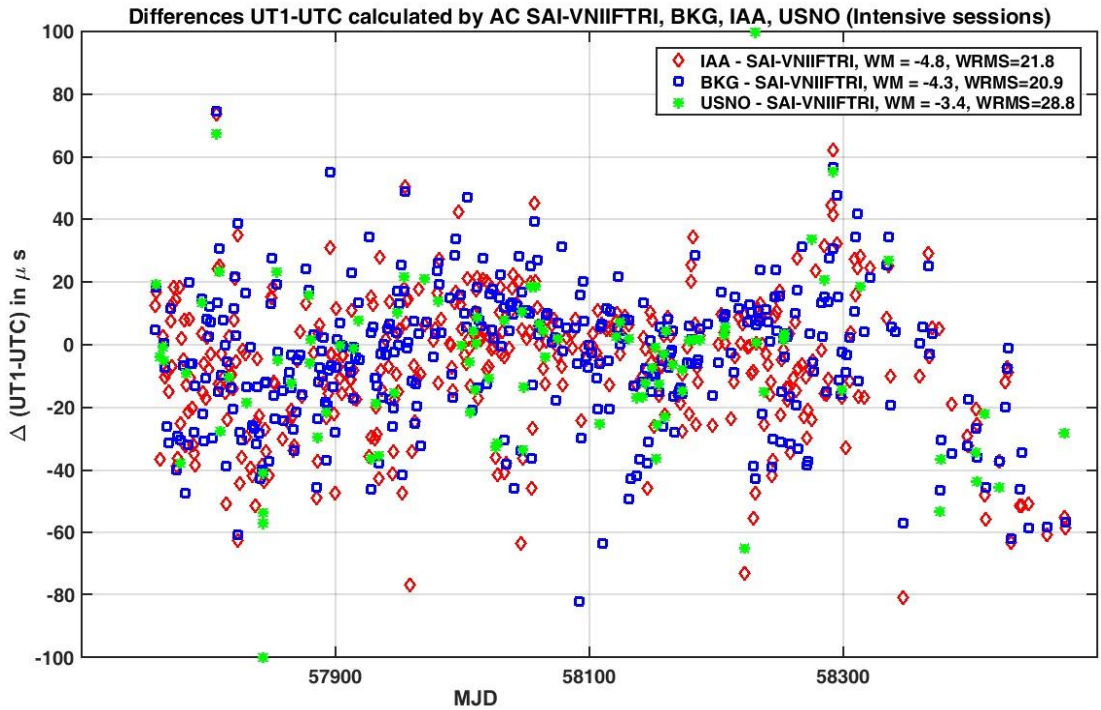

Fig. 2. AC BKG, IAA, USNO - AC SAI-VNIIFTRI UT1 differences from the solutions of Intensive sessions.

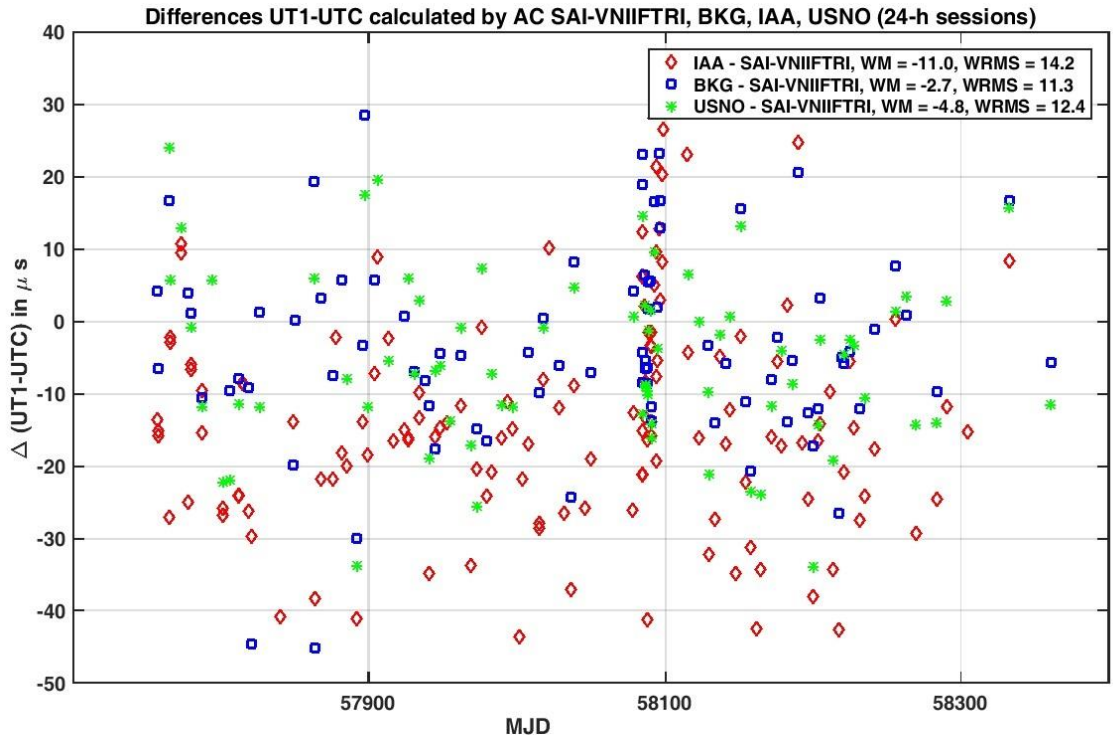

Fig. 3. AC BKG, IAA, USNO - AC SAI-VNIIFTRI UT1 differences from daily solutions.

VNIIFTRI plays role Main metrological center (MMC) of State Service for Time, Frequency and EOP evaluation (SSTF). It provides for customers include GLONASS information about time, frequency, national timescale UTC(SU) and Earth's orientation parameters EOP(SU) and does this reliable, precise and fast. All customers should be synchronized with few nanoseconds uncertainty. Now combining daily EOP(RU) are calculated in MMC SSTF by the processing of the nine independent individual EOP series provided by following four Russian analysis centers: MMC (VNIIFTRI), IAA (Institute of Applied Astronomy), IAC (Information-Analytical Center of Russian Space Agency) and SVOEVP (PNBS of Russian Space Agency). The experimental combination on base of the SINEX files processing for EOP and site coordinates evaluation is carried out in MMC too [Пасынок С.Л. и др., 2018].

# Positioning and Applications

Kosarev N.[1], Ustinov A. [1, 2]

[1] Siberian State University of Geosystems and Technologies, Novosibirsk, Russia
[2] JSC Institute Hydroproject, Moscow, Russia

During the last years, much attention has been paid to the monitoring of the structures, modernization algorithms processing of GNSS measurement, low-cost positioning, PPP technology, GNSS meteorology and use of terrestrial laser scanning technology for the different applications.

*Monitoring of the structures*.

In the paper [Канушин и др., 2017] the issue of the necessity of taking into account non-tidal changes in gravity during the deformation monitoring of hydraulic structures was touched upon. On a concrete engineering object, the results of calculating variations in the potential and variations in the normal height, caused by non-tidal variations in gravity, are presented. As such a test object, on which the results of calculations with the help of a specially developed software product were performed, the Sayano-Shushensky hydro-power unit performed. (Fig. 1)

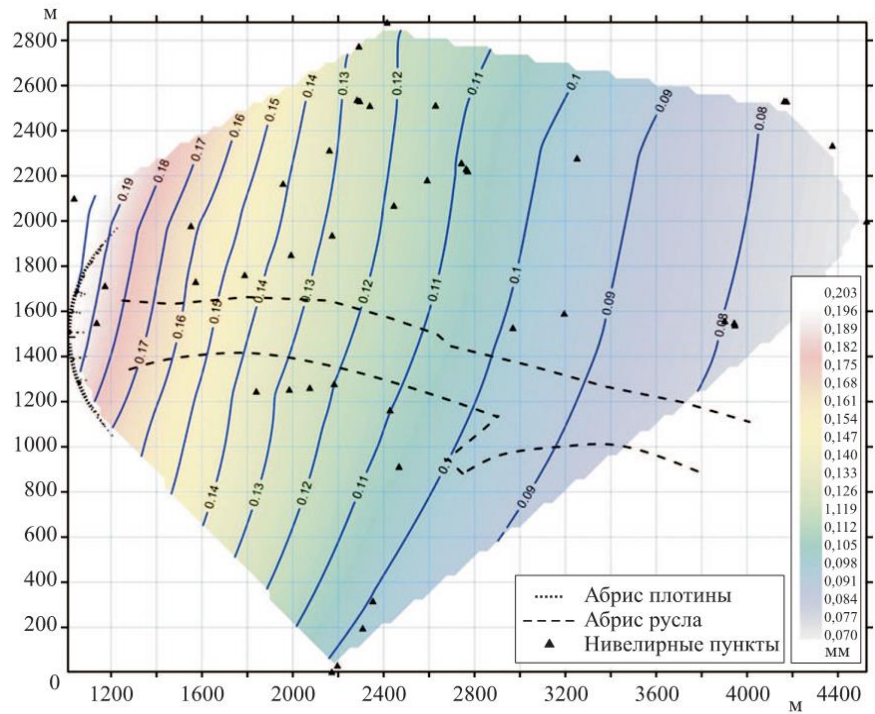

Fig. 1 Map of corrections to normal heights for the influence of variations in the gravitational field caused by a change in the level of the reservoir at 70 m

According to the results of the calculation, it is established that changes in the gravitational field caused by fluctuations in the level of the water masses of the reservoir make significant corrections to the results of geometric leveling, reaching an absolute value commensurate with the accuracy of the deformation monitoring of the hydraulic structure, which leads to the need to identify and correctly account for non-tidal variations of the gravitational field at deformation monitoring of hydraulic structures

Time series changes of coordinate increments obtained according to one-hour sessions of observation in the networks of local GNSS monitoring were investigated in the research [Устинов, Кафтан, 2016]. Revealed were the diurnal and semi-diurnal fluctuations of horizontal and vertical components of the vectors of reference lines with the length from 0.01 to 4.33 km and the amplitude up to 4 mm. The reasons of the identified fluctuations were tied to be found out. The possibility to model the fluctuations with a view to exclude them from the measurement results was shown. We can conclude that the application of digital filtering to exclude the fluctuation components from the measurement results is inadmissible.

In the paper [Устинов, 2018] considers the practical experience of application of technology of compensating injection (compensation grouting) for stabilization and rise of buildings and constructions is carried out, the analysis of results of compensation grouting given in domestic and foreign literary sources is carried out. The goals and objectives of the research work at the experimental site Zagorskaya PSP-2, made by specialists of JSC «Institute Hydroproject» in 2016–2017, are described. The scheme of work on the experimental plot and the description of technology of compensation grouting are given. The principles of organization of automated geodetic monitoring of displacements and deformations are shown on the example of the experimental site. The experience of geodetic observations of vertical movements of structures of the experimental site Zagorskaya PSP-2 in the process of compensation grouting is presented. The automated system of geodetic monitoring of movements of structures of the experimental site is described in detail (Fig. 2).

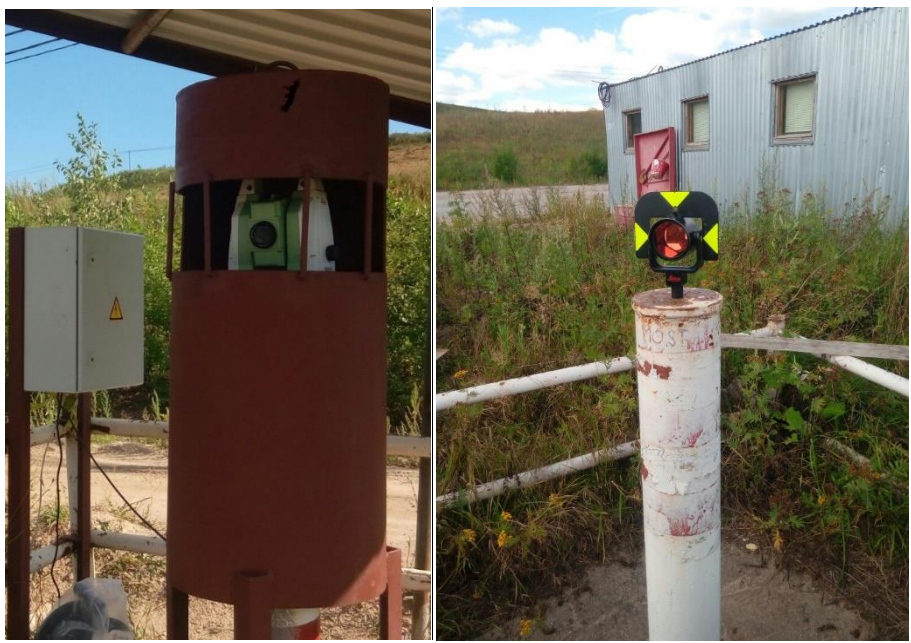

Fig. 2 Surveying equipment of experimental site

The results of observations of the vertical movements of the experimental site structures in the process of compensation grouting are presented. The results of automated monitoring of vertical displacements are compared with the results of class II leveling. According to the results of the comparison it was found that the accuracy of the automated determination of the altitude movements of the controlled points by deviations from the results of leveling of class II on average in cycles was ± 3.2 mm. Recommendations for improving the accuracy of automated systems of geodetic monitoring are given.

The technology of geodetic monitoring of hydropower structures during compensation grouting are described in [Устинов, Кафтан, 2019]. Recommendations on the optimal composition, configuration and functional structure of the automated geodetic monitoring systems based on robotic total stations are given. The process of data processing and analysis is shown on the example of the system of monitoring the rise of the facilities of the Zagorskaya PSP-2 experimental site. Special attention is paid to visualization of monitoring results.

*Low-cost positioning.*

The research single frequency GNSS receiver of OEM module NV08C-CSM Company NVS Technology AG in mode of RTK and static are presented in the article [Карпик и др., 2015а]. In these experiments dates of permanent base station located in Novosibirsk region are used (Fig. 3).

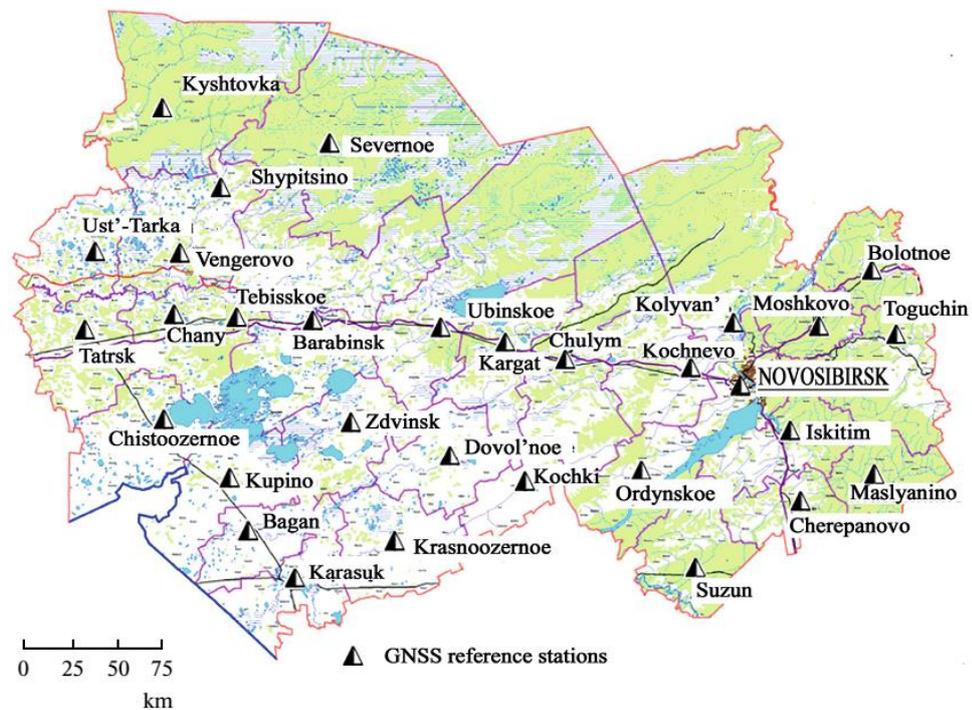

Fig. 3 The layout of the GNSS reference station network in the Novosibirsk region

Analyses experiments are made. Possibilities of using OEM-modules for application in geodesy and cartography tasks are proved.

The paper [Карпик и др, 2015б] deals with the version of transport accurate positioning navigation and information system using the budget GNSS equipment and the ground-based infrastructure of GLONASS. Navigation and information system bears complete nature; it consists of structurally complementary units and provides decimeter positioning accuracy.

The article [Karpik et al., 2016] describes a navigation and information system of precise transport positioning using low-cost GNSS equipment and the GLONASS ground infrastructure. The system consists of structurally complementary component parts and provides decimeter positioning accuracy (Fig. 4).

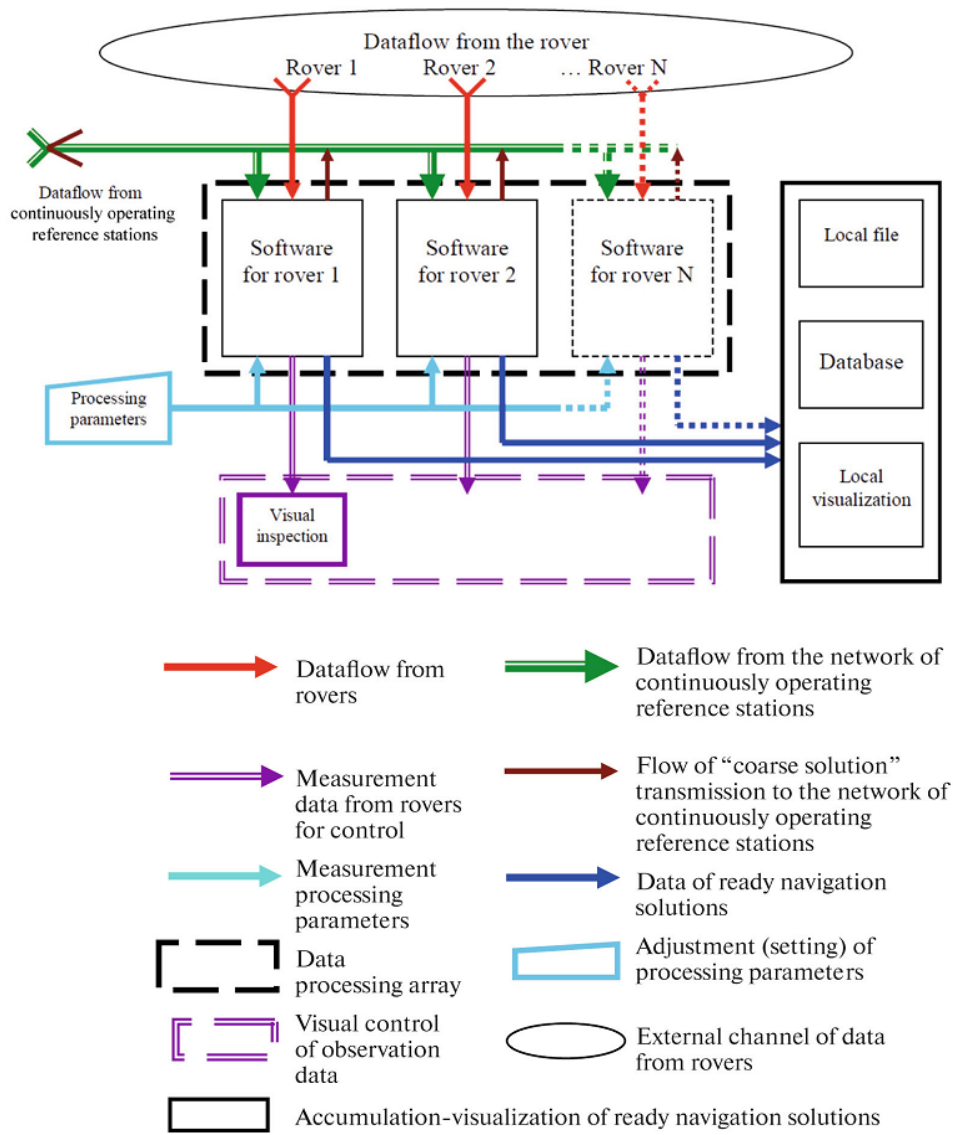

Fig. 4 Functional diagram of the hardware and software complex for dataflow control

*PPP technology.*

The estimation of the accuracy and precision of daily GNSS observations results using the Precise Point Positioning (PPP) method is presented in [Мельников, 2018]. The purpose of this estimation is substantiation of possibility of using the PPP method for the study of Earth's surface movements and deformations using geodetic methods. The estimation is based on the comparison of the daily static GNSS observations processing results by the PPP method with the results of the joint processing of daily static observations at the same point in the geodetic network using the traditional relative method for the double phase differences. Studied GNSS network consists of 13 permanent GNSS station which forms 30 baselines. For each baseline 364 differences of coordinates increments are calculated. Total selection is 10920 differences. The analysis of precision and accuracy of PPP in the context of previous studies was carried out. The analysis revealed that the achievable accuracy of the PPP method is comparable to the accuracy of the double difference method. Thus, the application of the PPP method to the study of geodynamic processes can be considered permissible only when the using the daily observations data and high-precision satellite ephemerides, satellite and station clocks data are involved.

In the research [Karpik, Lipatnikov, 2015] an approach aimed at providing compatibility and complementarity of real time kinematic (RTK) and precise point positioning (PPP), the well-known methods for high-precision positioning of mobile objects, using low-cost GLONASS/GPS equipment is proposed. Introduction of advanced modifications of these methods does not involve any changes in the existing ground infrastructure.

*GNSS meteorology.*

The dynamics of moisture content in the atmosphere on the territory of Leningrad and adjacent territories is investigated in research [Воротков и др., 2018] according to the GNSS database maintained in the Pulkovo Observatory areas. The average temperature of the atmosphere, necessary for the evaluation of PW, was calculated as data from the network of weather stations closest to the GNSS stations in the region and interpolated on the global database of atmospheric data NCEP/NCAR Reanalysis. Comparisons are made the obtained estimates of PW with those of the sounding of the atmosphere using water vapor radiometer and radio sounding of atmosphere. The maps with the time sweep the dynamics of the PW field for the study region. According to GNSS stations in the region with the most long-term observations evaluated the trend components of PW.

Data from selected permanent GNSS-stations of the countries surrounding the Gulf of Finland, separately for GPS and GLONASS observations are computed in research [Горшков, Щербакова, 2016] every day, the ranks of the location of the stations is using the strategy of PPP package GIPSY 6.3. Random and seasonal errors of these series were compared depending on the available three models of atmospheric and hydrological (groundwater) loads. Influence patterns of atmospheric pressures in a random attitude differ insignificant, while their seasonal components differ somewhat. The load models of seasonal variations of groundwater vary even more significantly. In General, GLONASS and GPS systems have no significant differences in the assessment of the positions of the stations selected for analysis.

In [Kalinnikov et al., 2018] results for the modeling of coordinate errors in the projected GNSS system of monitoring the Zagorskaya PSP-2 that are caused by local atmospheric irregularities in the surface layer are presented. It has been established that for mean meteorological conditions, these errors can reach 0.8 mm for elevation and 0.1 mm for horizontal position (Fig. 5). It is noted that actual instantaneous irregularities exceed mean values by an order of magnitude and can thus cause larger errors in coordinates.

a)
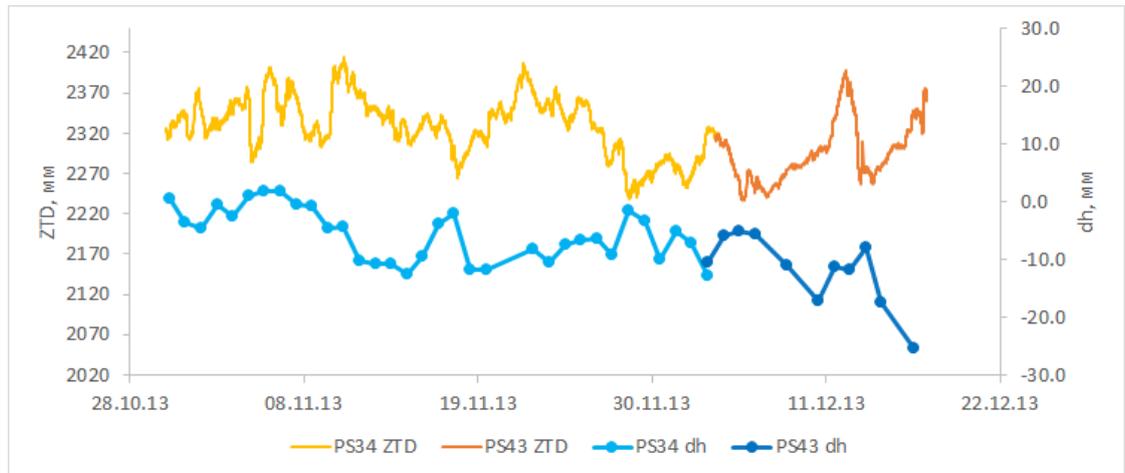

b)
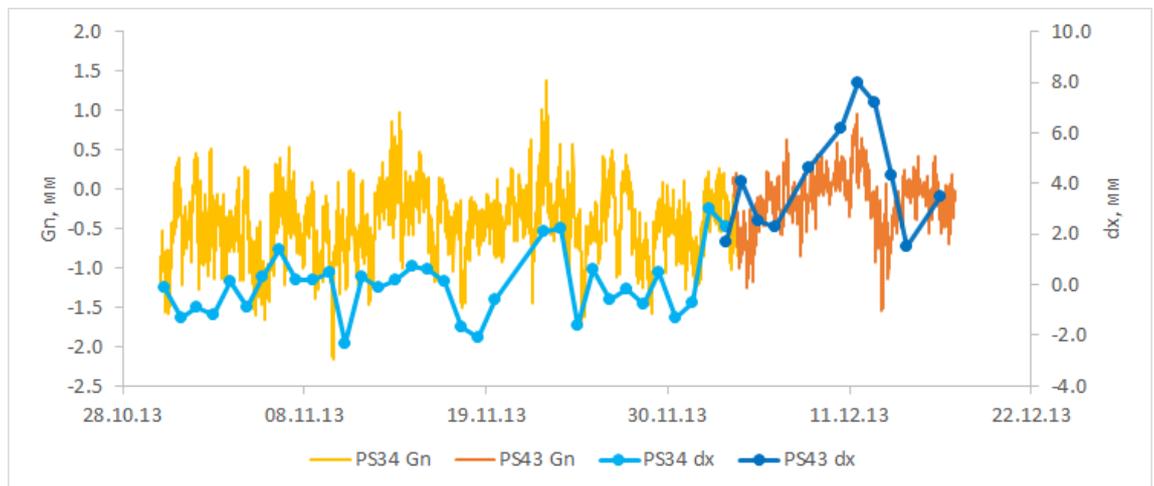

c)
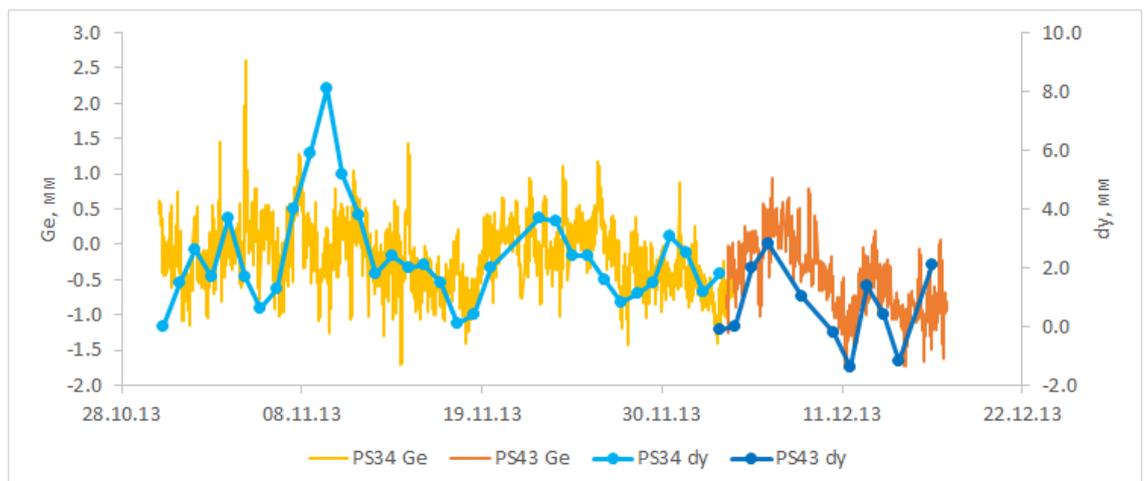

Fig. 5 Comparisons of ZTD series and height changes (a), the Northern gradient parameter and changes in the Northern coordinate (b), Eastern gradient parameter and Eastern coordinate changes (c)

*Modernization algorithms processing of GNSS measurement.*

The algorithm for taking into account the differential ionosphere delay effect is suggested in the article [Антонович и др., 2015]. The algorithm is applied for the control of two frequency phase GNSS measurements made by the receiver provided high stability frequency atomic oscillator. The algorithm approbation results made on the International GNSS Service station data are given. The obtained results are justified by the Global Ionosphere Model (GIM).

In the research [Безменов, 2018] an algorithm and a program for ultra-rapid Orbit/Clock estimation of GNSS (GLONASS & GPS) satellites based on observation data (in RINEX n/g/d format) received from 500 tracking stations of IGS network were developed. Test results of data processing for August 2017 were presented.

The article [Безменов и др., 2018] provides a detailed description of well-known method of smoothing the data of code measurements received from dual-frequency GNSS-receivers in the form of RINEX observation files (Receiver INdependent EXchange format). This method is widely used in processing programs for solving the applied problems in the field of geodynamics, time positioning and navigation service, in particular, in well-known Bernese GNSS Software. In the article the noise level evaluation for smoothed code data in the assumption that random components of the measurements are independent and centered values is presented. The method is based on solving the range of sub problems: 1) detection of receiver time jumps, 2) detection and removal of rough measurements (outliers) from data processing and 3) resolution of the carrier phase ambiguities. To solve these problems the techniques based on linear combinations from observation data are used. Conditions under which a receiver time jump is reliably detected against the background of measurement noise and possible jumps in phase ambiguities are formulated. So-called millisecond jumps are considered separately. For outlier detection a new algorithm minimizing a number of unreasonably rejected measurements is proposed. For detection of jumps in the so-called wide-lane phase ambiguity included in Melbourne-Wubbena combination a new algorithm based on formation of so-called clusters is proposed. The results of numerical experiments on comparison of the new algorithms with the algorithms applied in the Bernese GNSS Software are given.

Positions of GNSS stations can have non-tectonic variations in the form of the impulse noise which may last for several weeks in [Горшков, Щербакова, 2017] are shown. The analysis of GNSS data in the regional monitoring networks indicates that these position variations are often present simultaneously in the whole region as an in-phase disturbance. These spatial correlated variations in the GNSS networks are usually caused by the so-called common-mode error (CME). Various loading effects are assumed as the main reason of the CME. The influence of various meteorological parameters (the atmospheric pressure; and temperature and thickness of snow) on the appearance of these CME is searched in this study by the GNSS and meteorological data in the Baltic region. The GNSS data was processed by the GIPSY6.3 software tools; then all atmospheric and land water loading effects were taken into account. These CMEs are observed sometimes by the GNSS stations which are many hundred kilometers away from each other and often correlate with the atmospheric pressure and other meteorological parameters. This reveals an imperfection in the loading models used and/or inadequate amounts of meteorological stations included in the loading calculation. The CMEs appear in wintertime especially often being explained partly by snow level which is not used in the loading models.

The article [Дударев, 2018] considers phase path curvature influence of electromagnetic wave in the Earth's troposphere on the measured distance and the theory of taking account of this influence on the results of long-distance radio trajectory measurements of satellite. The article presents the necessary formulas for calculation of difference of electromagnetic wave phase path length from direct range line to the satelite. The article introduces mathematical models of non-query phase and impulse long-distance radio trajectory measurements of satellites, which are more appropriate to the real geometrical and physical conditions of measurement process and which take this difference into account on the correction level. Practical application of the suggested theoretical and methodological statements make it possible to increase the determination accuracy of state parameters of nonlinear dynamic system, consisting of several satellites and ground points. The parameters include spacecrafts' motion parameters, spatial coordinates of ground points, elements of relative orientation of different geodetic networks, Earth's rotation parameters and also some number of other geodetic and geodynamic parameters.

The experimental technology of usage of global navigation satellite systems for orientation and coordinate determination of observation tools of geomagnet stations and observatories are presented in [Кафтан и др., 2015]. The technology is experimentally tested at geomagnetic observatory St.-Petersburg. The prospects of the usage of permanent GNSS and geomagnet observation in a common observation network for fundamental geophysics research are discussed.

In the paper [Косарев и др., 2018] considers synthesized variants of the correlate and parametric versions of the least squares method (LSM)-optimization (adjustment) of geospatial data. Using synthesized variants of the correlate and parametric versions of the least squares method (LSM)-optimization was carried out equalization of GNSS measurement results by using software packages Trimble and CREDO, considering their correlation. As a test object was used a fragment of the satellite network of permanent base stations of the Novosibirsk Region consisting of 6 points: Novosibirsk (NSKW), Bolotnoe (BOLO), Iskitim (ISKT), Koluvan (KOLV), Kochenevo (KOCH), Suzun (SUZU). The correlated pair-data obtained as a result of such processing have been analyzed to test the zero hypotheses regarding insignificance of both individual differences and of their mean correlated value. In conclusion the paper proposes further ways of study to establish additional criteria for such analysis.

Multipath propagation of the navigation signal is one of the main sources of errors of high-precise GNSS-applications. There are a number of techniques to mitigate the effects of multipath at the GNSS- measurements on hardware and software level instrumentation. Thus from the perspective of the observation approach to minimize the effect of multipath in difficult conditions is not developed. The research [Куприянов и др., 2017] is dedicated to the development of an algorithm to calculate the coordinates of the navigation signal reflection point on an arbitrary surface, in order to study the impact of multipath effects on GNSS-measurements from a viewpoint of geometry and the development of methods of weakening the organization at the stage of observations.

The geodetic observations at geomagnetic observatories are used in the research [Kaftan, Krasnoperov, 2015] to orient reference directions in relation to a common coordinate grid. This problem is solved with the use of the measuring tools of global navigation satellite systems (GNSS). The results of experimental GNSS determinations at the St. Petersburg Geomagnetic Observatory, Russia, are presented (Fig. 6). Combination of magnetic and GNSS observations is proposed in order to reveal the cause-effect relationships between magnetic field variations and global geodynamic processes.

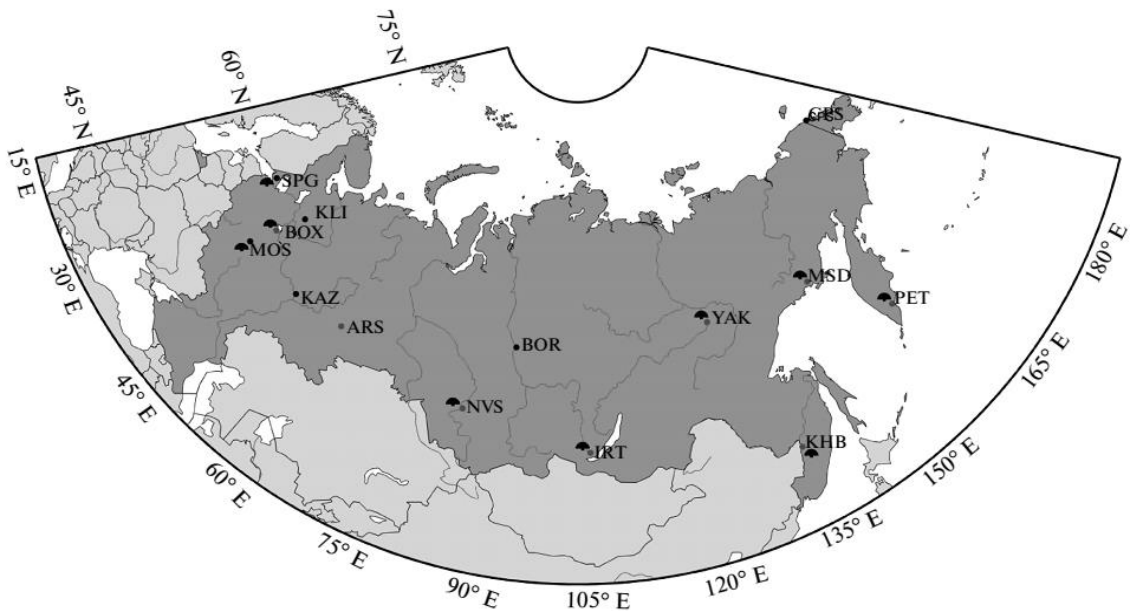

Fig. 6 The Russian INTERMAGNET segment and GNSS observation stations

Currently, one of the topical issues of improving GLONASS system is modernization of its uniformity measurement equipment, including RF

measurement equipment and electronic length measurement equipment. To this end, at the Spatial Reference Proving Ground of the Siberian State University of Geosystems and Technologies (SSUGT), the authors of the article [Karpik et al., 2018] carried out a successful experiment to measure a short GNSS baseline by receivers equipped with Chip Scale Atomic Clocks (CSACs) with instability of $10^{-11}$ showed that the mean deviation between the slant distance (D) measured using GNSS receivers connected to CSACs and their certified value varied in the range of 0.1–2.5 mm, with the average value of 0.9 mm (Fig. 7).

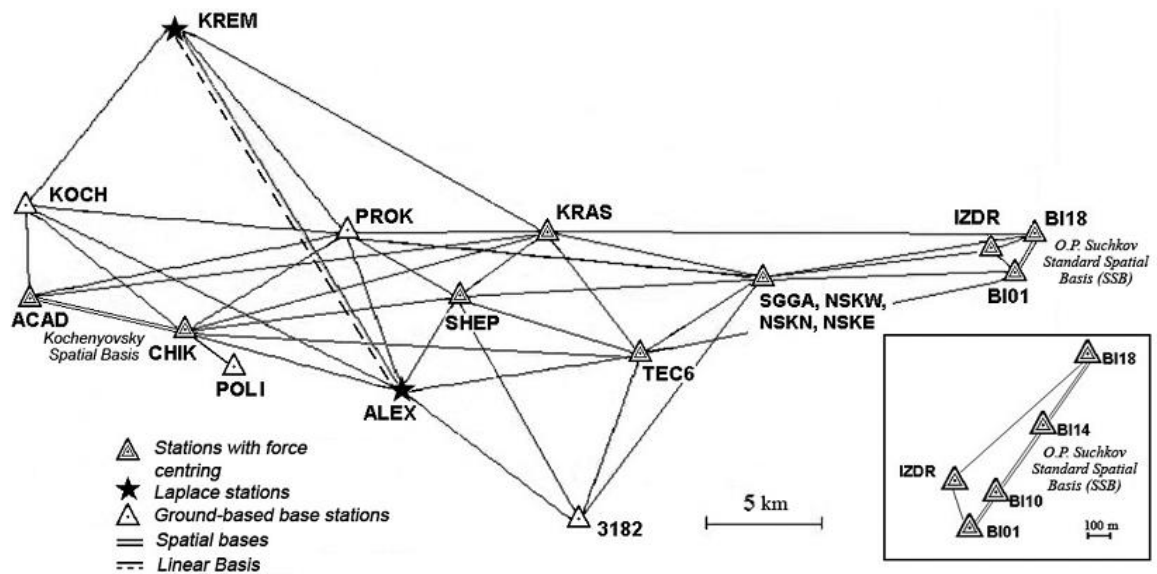

Fig. 7 GSPG network and O. P. Suchkov Standard Spatial Basis (SSB) of the SSUGT

The mean deviation obtained using GNSS geodetic receivers not connected to CSAC and their certified value made up 9.4 mm. The obtained experimental results suggest that substitution of quartz frequency generators with temperature compensation used in geodetic GNSS receivers for Chip Scale Atomic Clocks in any metrological or verification kit increases accuracy and reliability of short baselines measurements results, which highly perspective in view of development of techniques for creating reference baselines with a reproduction error of unit length of about 1 mm per 1 km. The above-mentioned experiment opens up new horizons for the use of Chip Scale Atomic Clocks in such fields of science as metrological support of geodetic equipment, geodesy, etc.

### *GNSS Orbits and clock*

The BERNESE 5.2 GNSS software is used for operative processing of GLONASS/GPS measurements in VNIIFTRI. The control script and some additional scripts were developed in VNIIFTRI for GLONASS/GPS operative orbits evaluation [Безменов 2018 a]. All scripts and programs are executed in the BERNESE 5.2 environments. BPE allows effective parallelization of estimations.

The experimental evaluations [Безменов 2018 b] are showed very good results for GLONASS clock corrections (Fig. 8).

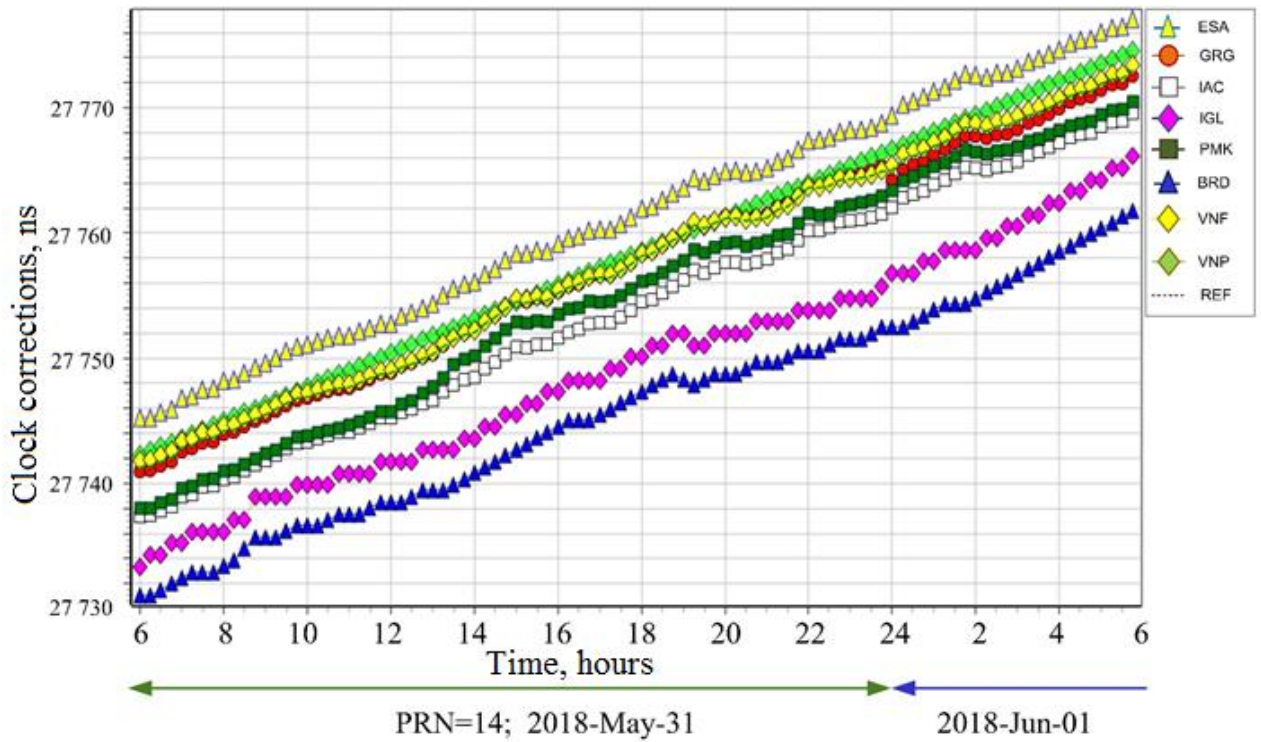

Fig. 8. GLONASS clock corrections in nanoseconds (VNF/VNP – results of processing/prediction of clock corrections in VNIIFTRI).

Because of the IGL GLONASS clock corrections values are alined to broadcast, the combined data were used as reference values for comparison (REF on Fig. 4). Combination software were developed in VNIIFTRI in 2015 [Безменов И.В., С.Л. Пасынок, 2015] and based on an realization of IGS orbits combination method. The orbits evaluation was not functioning as daily service in 2018.

*Terrestrial laser scanning technology.*

In [Ustinov et al., 2015] the necessity of using innovative technologies in the design of hydropower facilities is pointed out. The 3D modeling technology applied to design buildings, structures, and equipment of hydropower developments is described. Problems solved with the 3D modeling technology are detailed. Examples of using 3D models in reconstruction projects are presented. Prospects for further development of the technology are indicated.

A laser scanning and 3D modeling technology is presented in article [Ustinov, Bolodurin, 2016] for buildings, structures, and hydropower installation equipment. The tasks addressed using this three-dimensional technique are discussed in detail. Examples are presented and results are assessed of the use of the laser scanning technology within the scope of an integrated reconstruction project at the Nizhegorodskaya HPP. The advantages and prospects of laser scanning are noted.

# Common and Related Problems

Kaftan V.[1], Malkin Z.[2]

[1]Geophysical Center of the Russian Academy of Sciences, Moscow, Russia
[2]Pulkovo Observatory, Saint Petersburg, Russia

An overview of the many-year activity of the International VLBI Service for Geodesy and Astrometry (IVS) is made [Nothnagel et al., 2017]. The service regularly produces high-quality Earth orientation parameters (EOP) from observations conducted on global VLBI networks. These observations are also used to derive the International Terrestrial and Celestial Frames. The paper describes the current operations of the IVS, and an outlook is given of the next generation VLBI network Global Observing System (VGOS) based on the new VLBI2010 technology [Behrend et al., 2009].

A new method is proposed to divide a spherical surface into equal-area cells [Малкин, 2016а]. This method is based on dividing a sphere into several latitudinal bands of near-constant span with further division of each band into equal-area cells. It is simple in construction and provides more uniform latitude step between the latitudinal bands than other simple methods of equal-area tessellation of a spherical surface.

The history and results of using the Allan variance (AVAR) in geodesy and astrometry has been reviewed [Malkin, 2016]. In particular, there are issues with this method that need special consideration when analyzing these data. First, astronomical and geodetic time series usually consist of data points with unequal uncertainties. Therefore, one needs to apply data weighting during statistical analysis. Second, some sets of scalar time series naturally form multidimensional vector series, for example, 3D station position vector. The original AVAR definition does not allow to process unevenly weighted and multidimensional data. To overcome these deficiencies, two AVAR modifications has been proposed by the author.

Two astrometry conferences were held in the Pulkovo observatory in 2015 and 2018. Both conferences had a similar agenda with several sections with presentation related to the IAG activities. Section on ground based and space astrometry was in particular devoted to discussions of the reference systems frames. Most geodetic problems were discussed at the Section on Earth rotation and geodynamics. Problems of the data storage, dissimilation, and processing was discussed at a special section devoted to these topics. Historical section was also included in the program of both meetings. Group photo from these conference are presented in Figs. 1 and 2.

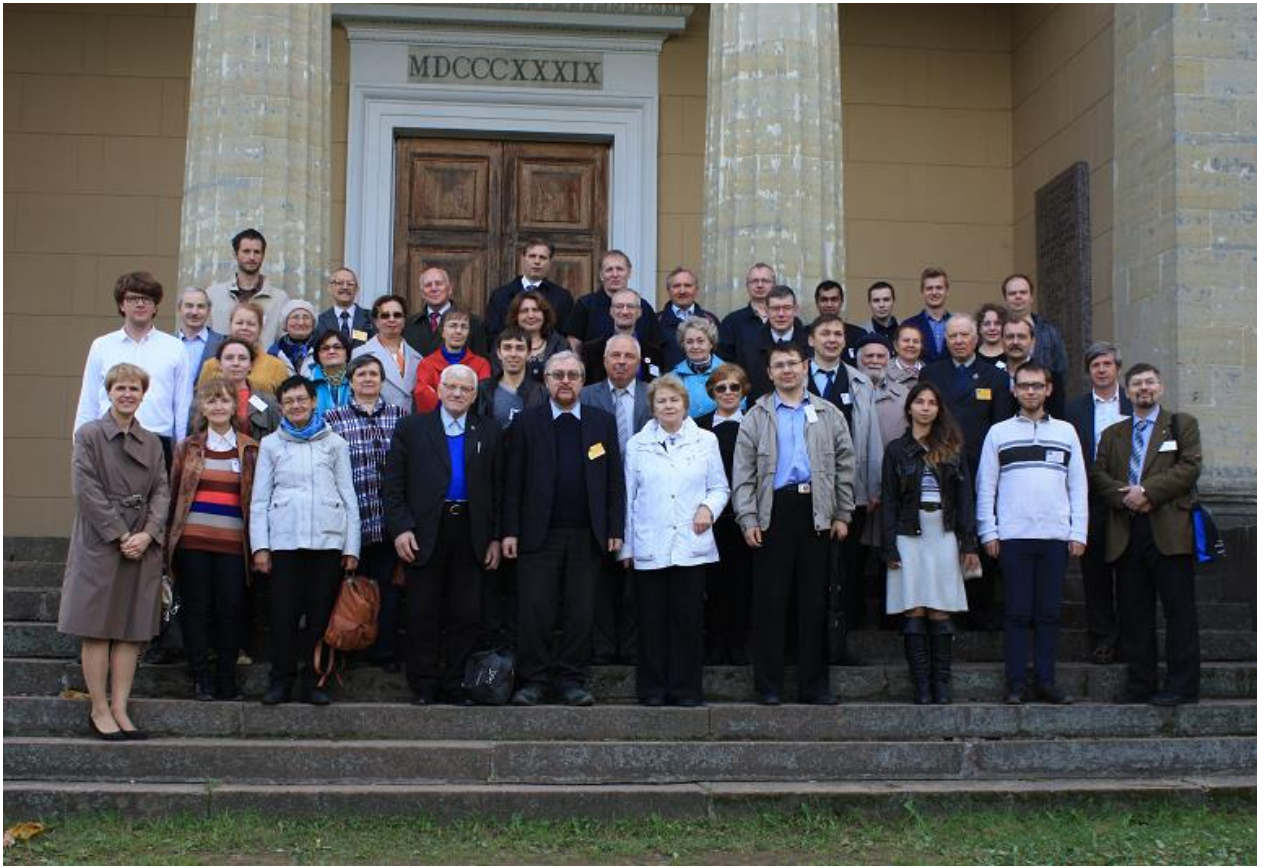

Fig. 1. Participants of the Pulkovo-2015 astrometry conference 21–25 Sept 2015.

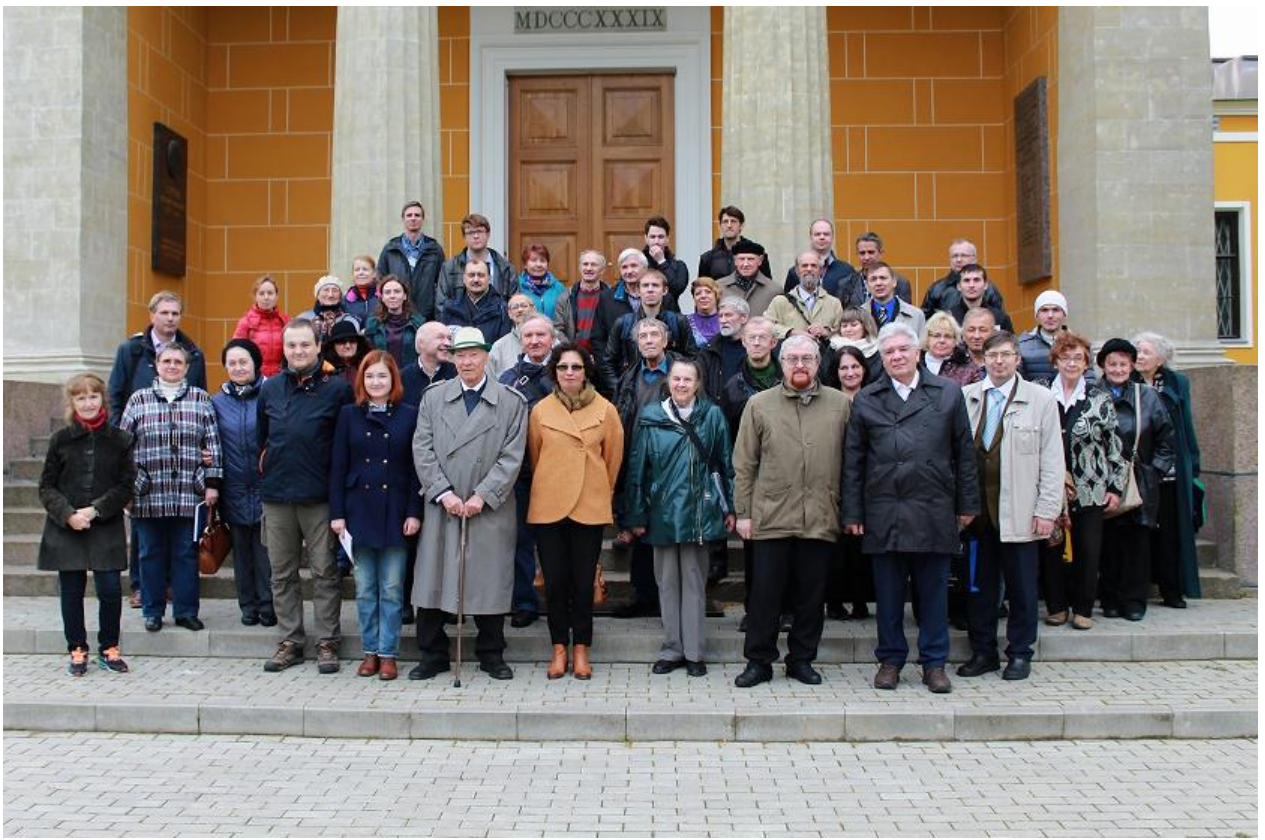

Fig. 2. Participants of the Pulkovo-2018 astrometry conference 1–5 October 2018.

Members of the Geodesy section of the National Geophysical Committee actively participated in the activity of the IAG Sub-Commission 1.4 "Interaction of Celestial and Terrestrial Reference Frames" [Malkin, Schuh, 2018].

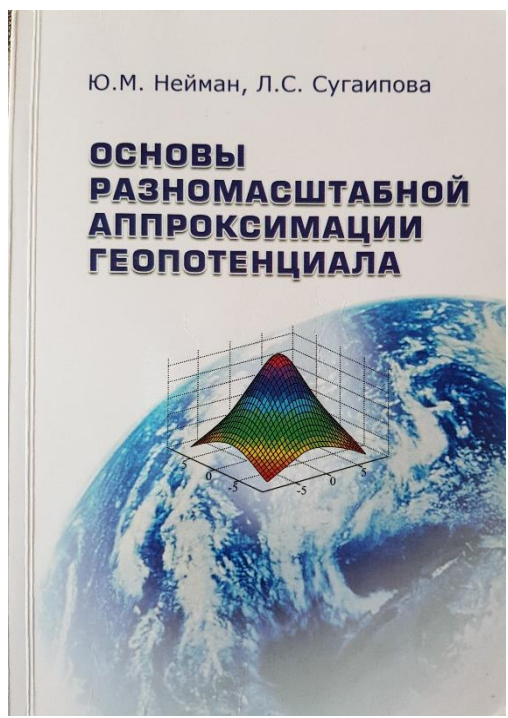

Fig. 3. Monography "Basics of multi-scale approximation of geopotential".

The monograph [Нейман, Сугаипова, 2016b] contains the key information on the multi-scale analysis and synthesis of signals in relation to gravitational field of the Earth. The monograph also includes recommendations for joint processing of heterogeneous measurements by the least squares method. Theoretical considerations are illustrated by examples from geodetic practice.

Scientific and technological paper collection issue was performed by the Federal Scientific Research Center of Geodesy, Cartography and SDI (TsNIIGAiK), Rosrrestr (Fig.4).

The following publications are devoted to the development of the theory and the development of new mathematical and numerical methods for solving geodetic problems [Безменов и др., 2018; Безменов, Блинов, 2015; Копейкин и др., 2016; Крылов, Яшкин, 2017; Липатников, 2016; Мазуров, 2015; 2016; 2017; Нейман, Сугаипова, 2015; 2016; 2019; Падве, Мазуров, 2017; Сурнин, 2015; 2018a; 2018b; Сурнин и др., 2015; Surnin, 2018a, 2018b].

New technological solutions, legal regulation issues, reviews of achievements in the field of geodesy and related areas are presented in [Побединский, Прусаков, 2016a, 2016b; Савиных, 2017; Цыба, 2018; ].

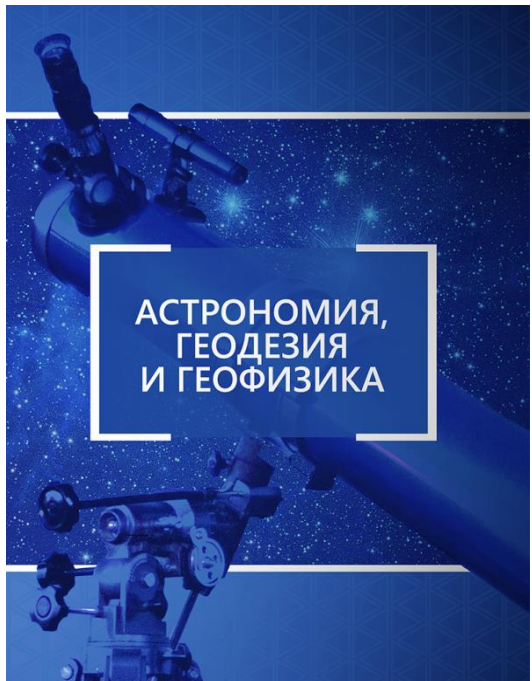

Fig. 4. Astronomy, Geodesy and Geophysics. Scientific and technological paper collection.

The combined time series of the Caspian Sea level using satellite altimetry data and classical observations was analyzed in order to identify relationships with global space-geophysical processes. [Kaftan et al., 2016, 2018].

Ionospheric disturbances in the vicinity of the Chelyabinsk meteoroid explosive disruption as inferred from dense GPS observations are revealed [Perevalova et al., 2015].